# Hot spots and flow structures around an isolated cuboid building subjected to surface warming: Large eddy simulations and wind tunnel measurements

Yi Zhao [1], Ruibin Li [1], Aytac Kubilay [2], Yongling Zhao [2], Xing Shi [3], Naiping Gao [1,*]

[1] *School of Mechanical Engineering, Tongji University, Shanghai, 201804, China*

[2] *Department of Mechanical and Process Engineering, ETH Zürich, Switzerland*

[3] *School of architecture and urban planning, Tongji University, Shanghai, 201804, China*

## Abstract

Urban warming is evident in numerous cities. On especially hot days, building surfaces warm up, leading to buoyancy-driven flows adjacent to these surfaces. The dynamics of the flow structures are largely influenced by the interplay between incoming wind and the buoyancy-driven flows. In this study, we used large eddy simulations and wind tunnel measurements to investigate the flow field around an isolated cubic building when different surfaces of the building are warmed. Under conditions of low wind speeds, ranging from 0.5 to 2 m/s, the surface temperatures of the scaled building were maintained between 20 and 95 °C. As the Richardson number ($Ri$) varied from 0 to 4.00, the flow, initially dominated by forced convection, shifted to being primarily steered by mixed convection. At low wind speeds and high $Ri$ values, the thermal effect led to noticeable changes in the reattachment and recirculation region lengths, with reductions of up to 48.3% in some cases. At pedestrian levels, thermally induced airflows often created localized hot spots, particularly around building corners and wall sections. This study offers insights into architectural designs that can potentially enhance wind-thermal comfort and improve pollutant dispersion around buildings.

*Keywords:* hot spots, flow structures, isolated building, wind tunnel tests, large eddy simulation

| Nomenclatures | | | |
|---|---|---|---|
| $C_s$ | Smagorinsky constant | $<U'W'>$ | vertical momentum flux |
| $H_{ref}$ | reference height (m) | *Abbreviations* | |
| $H_b$ | width of building (m) | ASM | Algebraic stress model |
| $k$ | turbulent kinetic energy(m$^2$/s$^2$) | IQ | index of quality |
| $L_x$ | computational domain length (m) | LES | large eddy simulation |
| $U$ | mean velocity magnitude (m/s) | RANS | Reynolds-Averaged Navier-Stokes |
| $U_x$ | mean *x* velocity magnitude (m/s) | Re | Reynolds number |
| $U_z$ | mean *z* velocity magnitude (m/s) | Ri | Richardson number |
| $U_{ref}$ | reference wind speed (m/s) | SGS | sub-grid scale |
| $U(h)$ | horizontal wind speed at height h (m/s) | UHI | urban heat island |
| $T_{ref}$ | reference temperature (°C) | *Greek symbols* | |
| $T_{ft}$ | average residence time (s) | $\alpha$ | power-law exponent |
| $\Delta t$ | time step | $\beta$ | coefficient of thermal expansion |
| $\mu_t$ | turbulent viscosity (N/(m$^2$·s)) | $\kappa$ | von Karman constant |
| $\tilde{u}_i$ | filtered mean velocity (m/s) | $\mu$ | dynamic viscosity |
| $\tilde{p}$ | filtered pressure (N/m) | $\rho$ | air density |

| $\widetilde{S_{ij}}$ | rate-of-strain tensor | $\Delta$ | grid scale |
| --- | --- | --- | --- |
| $\tau_{ij}$ | subgrid stress | $\delta$ | distance to the closest wall |

# 1. Introduction

With the drive of industrialization and economic development, urbanization continues to progress worldwide. As urbanization accelerates, cities face a common challenge known as the urban heat island (UHI) effect. It is expected that extreme hot and humid weather conditions will increase in the future, leading to even hotter and more oppressive cities compared to surrounding areas, thereby exacerbating the UHI effect [1]. Moreover, UHI effects contribute to increased energy consumption and heighten the risks associated with heat-related deaths [2, 3]. Solar radiation is a major factor contributing to UHI. In rural areas, there is a high vegetation coverage, where plants absorb solar radiation energy and release moisture through transpiration, leading to a decrease in surface temperature. However, in urban areas, factors such as buildings, human activities, and industrial emissions disrupt the absorption and release processes of solar radiation. Urban surfaces typically possess higher thermal conductivity and capacity, enabling them to absorb and store solar energy [4, 5]. Furthermore, high-rise buildings in cities impede airflows, leading to heat accumulation. Daytime temperatures and humidity are higher than those at nighttime, making the daytime crucial for studying urban thermal dynamics. During the daytime, the surfaces of the building are exposed to solar radiation, causing surface temperatures to rise, and the thermal effect simultaneously alters wind speed and airflow patterns. The ventilation and dispersion of pollutants in urban areas are significantly impacted by the structure of the flow field [6]. Factors such as airflow velocity, direction, and distribution play a vital role in the design of urban ventilation systems. Additionally, vortices and turbulence in the flow field structure can have an impact on the dispersion and mixing of pollutants.

In comparison to other building configurations, the isolated building has a simpler structure but exhibits significant airflow characteristics in its surroundings. The absence of surrounding buildings allows the airflow to move more freely, bypassing the building and creating specific aerodynamic effects in the vicinity. Reynolds-Averaged Navier-Stokes (RANS) and Large Eddy Simulation (LES) are the most commonly used turbulence models in building aerodynamics. While RANS models are widely employed in various applications, LES is better suited for simulating complex flow phenomena around buildings, as it has the capability to capture transient features of the flow field. [7, 8]. One drawback of utilizing LES is the higher computational cost in terms of time [9]. Nevertheless, this limitation can be addressed by leveraging the rapid advancements in computing resources and the continuous improvements in computational power [10]. Additionally, the study by Krajnovic and Davidson [11] indicates that obtaining accurate LES results in coarse grids is highly feasible, making it possible to reduce the computational cost of LES. Nevertheless, a unified best practice for LES applications is still lacking, and the validation of LES results requires in-situ observations or wind tunnel tests [12-14]. Murakami et al. [15] were the first to apply LES to predict the flow field around a cubic building in the atmospheric boundary layer. Subsequently, they compared turbulent models such as *k-ε*, Algebraic stress model (ASM), and LES with wind tunnel tests to assess the differences in predicting the flow field around a cubic building [16]. The LES results showed excellent agreement with the wind tunnel test results, successfully reproducing the velocity distribution in the reattachment region on the roof and the recirculation region on the leeward side of the building. Similarly, Richards and Norris [17] utilized LES to simulate the unsteady flow around a cubic building and effectively reproduced the observed flow patterns on a full-scale model. Okaze et al. [18] obtained high-quality measurements of fluctuating velocity and mean velocity around an isolated building at a scale of 1:1:2 using wind tunnel tests and LES. Their study focused on comparing the effects of grid refinement, sub-grid scale (SGS) models, and discretization schemes for the convective term on the computed results. Liu et al. [19] investigated the impact of different roof shapes on the airflow around the isolated building using LES and wind tunnel tests. Furthermore, LES has been employed to study the airflow and pollutant dispersion processes around the cubic building under different thermal stratification conditions [20], as well as within and above the street canyon with different heated surfaces [21-23].

However, based on the literature review conducted so far, most wind tunnel tests and numerical simulations have focused on the flow patterns around the isolated building under isothermal conditions. Due to the numerous challenges associated with reproducing solar radiation in wind tunnels and water tunnels [24], only a limited amount of study has been dedicated to examining the flow characteristics around the isolated building under the influence of thermal effects through wind tunnel tests [25, 26] and within urban street canyons [27-29]. A comprehensive investigation of flow characteristics around the isolated building under non-isothermal conditions is currently lacking. Therefore, a thorough analysis of wind tunnel test data and validated numerical simulation results of wind thermal fields around buildings with thermal effects is a meaningful endeavor.

The aim of this study is to comprehensively investigate the flow field around an isolated building, considering both forced-convective and buoyancy-driven flows. Using a combination of LES and wind tunnel tests, we reveal the flow field, airflow paths, vertical momentum flux, and turbulence kinetic energy distribution under varying heating conditions. We use a 1:1:2 cuboid; during wind tunnel tests, a heating film is applied to elevate the surface temperature of the building. Subsequently, the windward wall, leeward wall, and all building walls are heated ranging from 20 °C to 95 °C at varying incoming flow velocities, allowing for comparisons with the flow field around an isothermal building. The Richardson number ($Ri$) is employed to assess the dominance of natural convection and forced convection around the building, ranging from 0 to 4 in our study. Furthermore, we track the variation in the scale of the vortex region and the trajectory of the vortex center across investigated cases, ensuring detailed flow field characterization.

The structure of this paper is organized as follows: Section 2 provides details of the LES and outlines the cases explored in this study. Section 3 elucidates the setup of the wind tunnel test, along with a comparison between the results obtained from the wind tunnel tests and LES. Section 4 conducts a holistic analysis of flow field characteristics around the building, taking into account various intensities of buoyancy-driven flows. A comparative analysis is conducted across four distinct conditions: absence of heating, heating applied to the windward wall, heating on the leeward wall, and heating encompassing all walls. Discussions and conclusions are presented in Section 5 and Section 6, respectively.

## 2. Outline of CFD simulation

### 2.1 Governing Equations

The numerical simulations in this study were performed using the CFD software Ansys Fluent that has been adopted in recent studies [30]. In LES, the computation resolves the large-scale fluctuations, while the influence of small-scale fluctuations on the large scales is modeled. LES employs filtering methods to separate the small-scale fluctuations in turbulence, and its governing equations involve the application of filtering operations to the instantaneous motion of the Navier-Stokes equations. The filtered governing equations are expressed as follows [31]:

$$\frac{\partial \rho \widetilde{u}_i}{\partial x_i} = 0 \tag{1}$$

$$\frac{\partial}{\partial t}(\rho \widetilde{u}_i) + \frac{\partial}{\partial x_j}(\rho \widetilde{u}_i \widetilde{u}_j) = \frac{\partial}{\partial x_j}\left(\mu \frac{\partial \widetilde{u}_i}{\partial x_j}\right) - \frac{\partial \widetilde{p}}{\partial x_i} - \frac{\partial \tau_{ij}}{\partial x_j} \tag{2}$$

where $\widetilde{u}_i$ represents the filtered mean velocity, $\widetilde{p}$ are the filtered pressure, $\rho$ are the air density, and $\mu$ are the dynamic viscosity.

$\tau_{ij}$ represents the subgrid stress, which represents the momentum transport between the filtered small-scale pulsations and the resolved-scale turbulence. It is modeled as follows [32]:

$$\tau_{ij} = -2\mu_t \widetilde{S_{ij}} + \frac{1}{3}\tau_{kk}\delta_{ij} \tag{3}$$

$$\widetilde{S_{ij}} = \frac{1}{2}\left(\frac{\partial \widetilde{u}_i}{\partial x_j} + \frac{\partial \widetilde{u}_j}{\partial x_i}\right) \tag{4}$$

where $\widetilde{S_{ij}}$ is the rate-of-strain tensor for the resolved scale, and $\mu_t$ is the subgrid-scale turbulent viscosity.

The subgrid-scale stresses generated by filtration need to be modelled. In the Smagorinsky-Lilly model, the eddy-viscosity is modeled by [32]:

$$\mu_t = \rho L_s \sqrt{2\widetilde{S_{ij}}\widetilde{S_{ij}}} \tag{5}$$

$$L_S = min(\kappa \delta, C_s \Delta) \tag{6}$$

where $\kappa$ is the von Kármán constant, $\delta$ is the distance to the closest wall, $\Delta$ is the local grid scale and $C_s$ is the Smagorinsky constant.

2.2 Computational domain and grid

As shown in Fig. 1, a rectangular computational domain is established around the isolated building with dimensions of 31$H_b$(x) ×21$H_b$(y) × 12$H_b$(z), where $H_b$ represents the width of the isolated building. The dimensions of the isolated building model are 0.2 m (length) × 0.2 m (width) × 0.4 m (height), while the prototype dimensions are 10 m (length) × 10 m (width) × 20 m. This results in a scale ratio of 1:50. The computational domain dimensions follow the recommendations of COST [33]. The origin of the coordinates is positioned at the leading edge of the isolated building on the ground, as indicated in Fig.1. The blockage ratio in the direction of the incoming flow is kept below the recommended 3% by the AIJ within the computational domain [34].

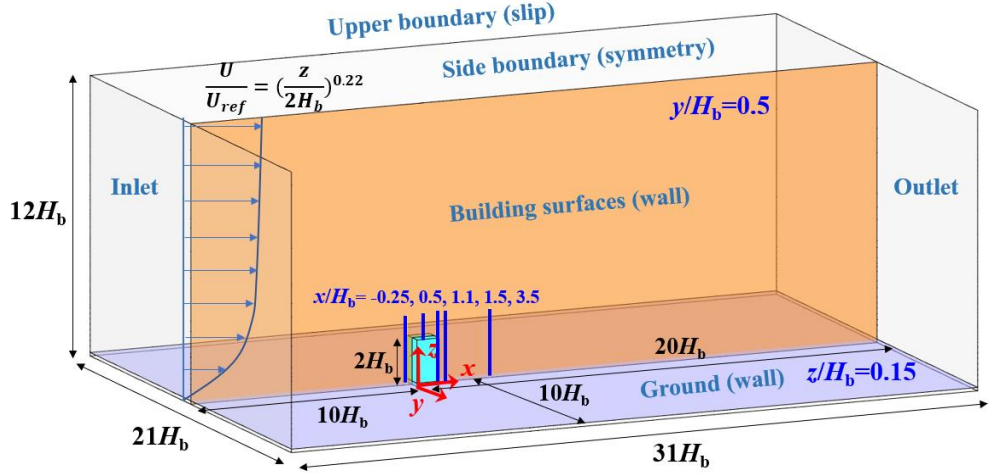

**Fig. 1.** Diagram of computational domain and the location of key lines and slices.

In this study, three distinct structured grids (coarse, medium, and fine) were generated using ANSYS ICEM CFD 19.0. Their respective cell numbers are approximately 3 million, 7 million, and 16 million. Fig. 2 depicts the computational grids used in this study. For the medium grid, the width of the isolated building was discretized into 30 evenly spaced grid cells, while the coarse and fine grids were discretized into 23 and 39 cells, respectively. The grid spacing in the building height direction was the same as in the width direction. The grid expansion ratio for all cases was set at 1.1, which is lower than the recommended value of 1.3 [34]. This approach allows the grid cells to expand gradually as they move away from the building surfaces, leading to a reduction in the total number of grid cells and computational costs.

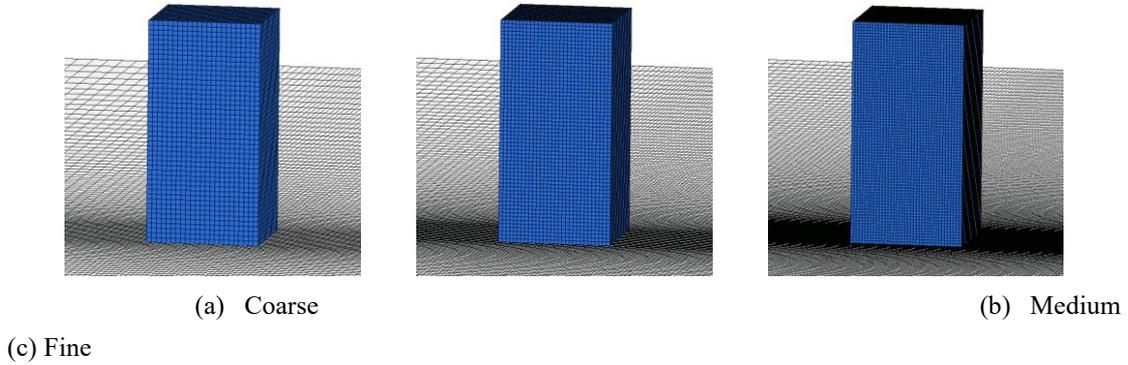

(a) Coarse  (b) Medium

(c) Fine

**Fig. 2.** Grid arrangement around the building.

The Index of Quality (IQ) is employed to determine whether the mesh quality of the LES satisfies the solution requirements. IQ is defined as the ratio of resolved kinetic energy to the total kinetic energy. Celik et al. [35] and Shirzadi et al. [36] previously utilized IQ to estimate the ratio of resolved-scale kinetic energy to total kinetic energy. Fig. 3 illustrates the normalized, time-averaged $U_x$ profiles, representing streamwise component of velocity, derived from LES using three different grid settings. The predicted results for the medium and fine grids are highly similar, while the coarse grid shows slightly higher predicted values on the leeward side of the building compared to the medium and fine grids. The mean values of IQ for different measurement lines are presented in Table 1. The mean values of IQ for all measurement lines are above 85%, surpassing the recommended threshold for LES. Considering both grid performance and computational costs, the medium grid is selected for the subsequent LES in this study.

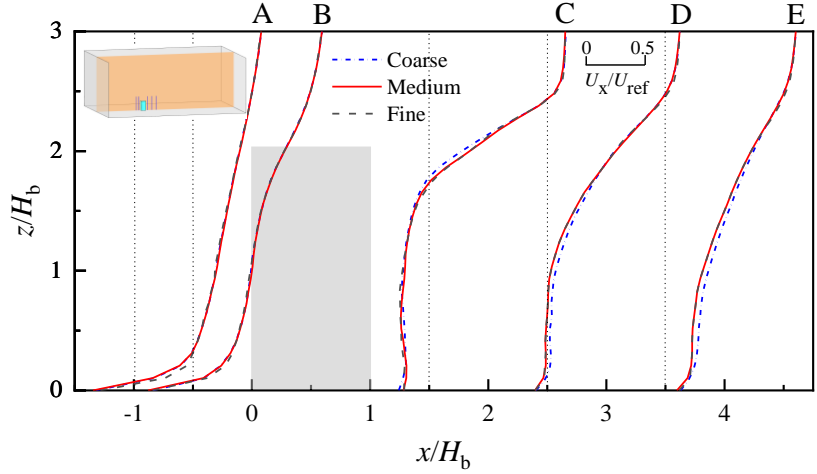

**Fig. 3.** Comparison of vertical profiles of normalized time-averaged $U_x$ for three different grid settings in LES.

**Table1**
Five measurement lines corresponding to the mean values of IQ for three different cell numbers.

| IQ (%) | A | B | C | D | E |
| --- | --- | --- | --- | --- | --- |
| Coarse | 87 | 89 | 89 | 88 | 86 |
| Medium | 89 | 92 | 92 | 91 | 92 |
| Fine | 90 | 93 | 90 | 94 | 92 |

## 2.3 Solver settings

In this study, the SIMPLE solver is utilized for pressure-velocity coupling. A second-order scheme is employed for pressure interpolation, while a bounded central differencing scheme is used for spatial discretization of the momentum equation. The transient formulation utilizes a bounded second-order implicit scheme. The Smagorinsky-Lilly model is utilized, and the value of the Smagorinsky constant ($C_s$) is determined based on previous LES studies on flow around buildings which suggested a value of $C_s$=0.12 [7, 37-39]. Therefore, in this study, $C_s$=0.12 is adopted to investigate the flow field around the isolated building. And, the turbulent Schmidt number $\mu_t$ was set to 0.5 [40]. The velocity fluctuations at the inlet boundary are computed using the Vortex method [36, 41]. The Courant numbers for all grids are maintained below 0.9, and the range of time steps $\Delta t$ corresponding to different cases is 0.003 to 0.009.

In LES, selecting an appropriate time step and ensuring a sufficient simulation duration are crucial. Initially, a steady Reynolds-averaged Navier-Stokes (RANS) model is conducted as a reference, running until statistically stable solution is achieved. It is worth noting that both the steady RANS and LES simulations employ the same grid and computational domain. Following this, time interpolation is applied to the steady RANS solution to derive instantaneous velocity fields at various time steps. For the RANS method, the SIMPLE method resolves the pressure-velocity coupling, while the the standard $k$-$\varepsilon$ model represents turbulence characteristics. The duration of LES can be determined by estimating the average residence time, $T_{ft}$, within the computational domain, calculated as $T_{ft}=L_x/U_{ref}$ (where $L_x$ represents the length of the computational domain). In this study, the LES spanned 22 $T_{ft}$, surpassing the sampling duration suggested by Gousseau [31].

## 2.4 Case description

The LES analysis examines the evolving patterns of the flow field surrounding an isolated building under the combined effects of varying surface heating temperatures (45 °C, 75 °C, 95 °C, with a reference temperature $T_{ref}$ set at 20 °C), distinct incoming flow velocities (0.5 m/s, 1 m/s, 2 m/s) at building height, and specific heating configurations (isothermal, windward wall heating, leeward wall heating, and all-wall heating). The incoming flow direction aligns perpendicularly to the windward wall of the isolated building. The interplay between forced convection and natural convection is characterized by the Richardson number, $Ri$ [24], which ranges from 0 to 4 in this study. The wall temperature and incoming flow velocities correspond to the Reynolds number ($Re$) and $Ri$, as detailed in Table 2. Smaller $Ri$ values signify forced convection, whereas larger $Ri$ values indicate mixed convection. For instance, "LEE-0.5-45" denotes heating of the leeward wall to 45 °C with an incoming flow velocity of 0.5 m/s at the $H_{ref}$. "ISO-0.5" indicates an absence of heating with an incoming flow velocity of 0.5 m/s at the $H_{ref}$. Likewise, "WIN" represents heating of the windward wall, while "ALL" signifies heating across all walls.

$$Ri = \frac{\beta g H_{ref}(T_w - T_{ref})}{U_{ref}^2} \quad (7)$$

where $U_{ref}$ is the reference velocity, $\beta$ is the coefficient of thermal expansion, $T_{ref}$ is the ambient temperature, $T_w$ is the wall temperature and $g$ is the gravitational acceleration, $H_{ref}$ is the height of the building, 0.4m.

**Table 2**
Surface temperatures and incoming flow velocities with corresponding $Re$ and $Ri$ numbers.

|  | 20 °C | 45 °C | 75 °C | 95 °C |
|---|---|---|---|---|
| 0.5 m/s ($Re$=7643) | 0 | 1.34 | 2.94 | 4.00 |
| 1.0 m/s ($Re$=25478) | 0 | 0.33 | 0.74 | 1.00 |
| 2.0 m/s ($Re$=50955) | 0 | 0.08 | 0.18 | 0.25 |

## 3. Wind tunnel tests and validation

### 3.1 Wind tunnel test settings

The measurements were conducted using the TJ-1 atmospheric wind tunnel. The TJ-1 wind tunnel features a cross-section measuring 1.8 m × 1.8 m, providing a spacious testing area. The test section itself spans a length of 12 meters, allowing for airflow simulation and analysis. To investigate the airflow characteristics over urban terrain, a series of wind tunnel tests were conducted. The purpose of these experiments was to simulate the complex wind patterns that occur in cities, taking into account factors such as the roughness of surfaces and the presence of buildings. To generate turbulence and increase surface roughness, specific elements with spires and rough elements were incorporated into the wind tunnel setup. The mean wind profile, representing the distribution of wind speeds at different heights, was modeled using a power-law function with a power exponent $\alpha$ of 0.22 (Eq.8). During the wind tunnel tests, the mean wind speeds at various heights of the building model were recorded. The turbulence intensity, an indicator of the level of turbulence in the airflow, was estimated to be around 10% at the front of the building.

$$\frac{U(h)}{U_{ref}} = \left(\frac{h}{H_{ref}}\right)^\alpha \quad (8)$$

Where $\alpha$ represents the power exponent, $U(h)$ is the horizontal wind speed at height h.

The handle of the hot-wire anemometer used in the wind tunnel test is the Testo 400,

manufactured in Germany. It is paired with a digital probe connected to the Testo 400 via Bluetooth, capable of simultaneously measuring both wind speed and temperature. The wind speed measurement range spans from 0 to 50 m/s. When the wind speed is below 20 m/s, the measurement accuracy is ±0.03 m/s + 4%, with a resolution of 0.1 m/s. As for temperature, it can be measured within a range of -20 to 70°C. The measurement accuracy for temperatures between 0 and 70°C is ±0.5°C, and for temperatures between -20 and 0°C, it is ±0.8°C, with a resolution of 0.1°C. Prior to each test, the hot-wire anemometer undergoes calibration, and each sampling session lasts for 5 min. The 3D-Cobra probe is used to measure the turbulence characteristics of airflow. This instrument is manufactured by Turbulent Flow Instrumentation Pty Ltd, an Australian company specializing in the development of instruments for the measurement of turbulent flow fields. The 3D-Cobra probe features a linear frequency response range spanning from 0 Hz to over 2 kHz and can measure wind speeds ranging from 2 m/s to 100 m/s with a precision of ±0.5 m/s.

Comparing the profiles of mean wind speed and turbulence intensity obtained from experiments and LES, the results are shown in Fig. 4, providing a visual and analytical representation of experimental and simulation data. In the wind tunnel test, the isolated building model was placed on the wind tunnel turntable, and the measured mean velocity and turbulence intensity profile (measured at $x/H_b = -1$), as well as the simulated profile from numerical simulation, were depicted in Fig. 4(d) and 4(e). The velocity profiles and turbulence intensity levels in the experimental setup closely align with the settings used in LES. The physical model used in the wind tunnel tests was constructed using wooden panels with dimensions of 0.2 m (length)×0.2 m (width)×0.4 m (height). Heating films were applied and controlled by digital temperature controllers in order to set the surface temperature of the building model, ensuring a constant and controlled temperature distribution throughout the test. This thermal manipulation helped capture the influence of temperature gradients on the airflow over urban surfaces.

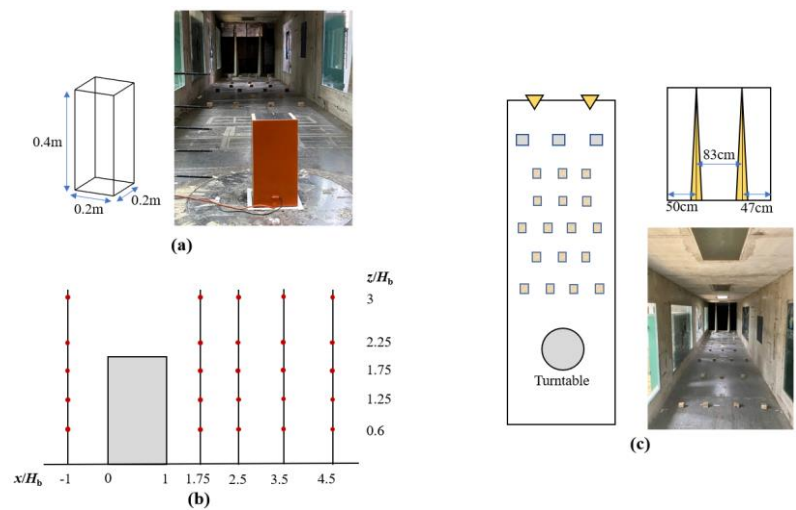

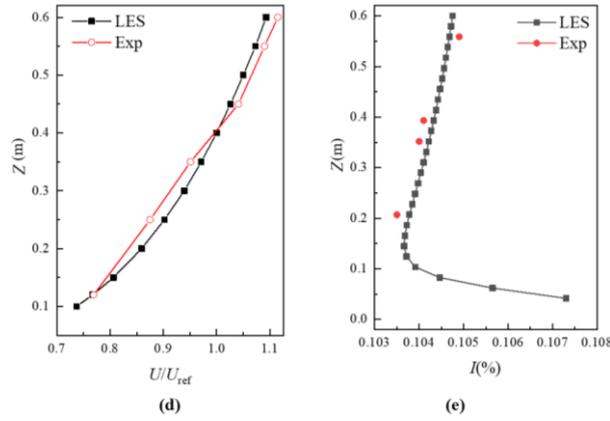

**Fig. 4.** The diagram of (a) dimensions and configuration of the model building, (b) location of measurement points, (c) arrangement of spires and roughness elements, (d) the magnitude of mean velocity profile of the incoming flow and (e) the turbulence intensity of the incoming flow.

3.2 LES validated by wind tunnel test results

The LES results were validated by comparing them with the measured values obtained from wind tunnel tests conducted at TJ-1 wind tunnel. The profiles of the normalized mean velocity ($U/U_{ref}$) and temperature ($T$) near the building are depicted in Fig. 5. These profiles were obtained along vertical lines by selecting the wind speed and temperature results at $x/H_b$ = -1, 1.75, 2.5, and 3.5 from the wind tunnel tests as indicated in Fig. 4(b). While this paper involves a variety of cases, the differences in the results between the two methods, i.e., numerical and experimental, are similar. Therefore, in Fig.5(a), only the ISO-0.5 is selected to compare the normalized time-averaged velocity. On the windward side measurement line at $x/H_b$ = -1, the experimental results for the normalized mean velocity were slightly higher than the simulated results. However, the remaining measurement lines demonstrate a high level of consistency between the two methods. Upstream of the building, the profiles are significantly affected by the inlet conditions, and the LES and experimental settings are not perfectly aligned. Additionally, the flow development in numerical simulations follows a distinct pattern. As we move downstream, the influence of the building becomes more pronounced, and LES excels in accurately capturing this phenomenon.

Fig. 5(b) presents a comparison of the mean temperatures for the WIN-2.0-45 case. It's important to note that the mean temperature obtained in the experiment represents a five-minute average. Regarding mean temperature, the results from LES are slightly lower than those from the experiments. This distinction is most prominent along the measurement line at x/Hb=-1, which corresponds to the windward side of the building. This discrepancy may be attributed to a temperature hysteresis of approximately 1°C that occurs when the heating is employed on the windward side. It is important to highlight that the disparities between the two methods at equivalent building height along various measurement lines exhibit a consistent pattern. For example, at the location where $z$=0.6 m, the outcomes derived from both methods closely resemble each other. Overall, with only a few isolated data points as exceptions, the results from the wind tunnel tests closely correspond to those obtained from the LES.

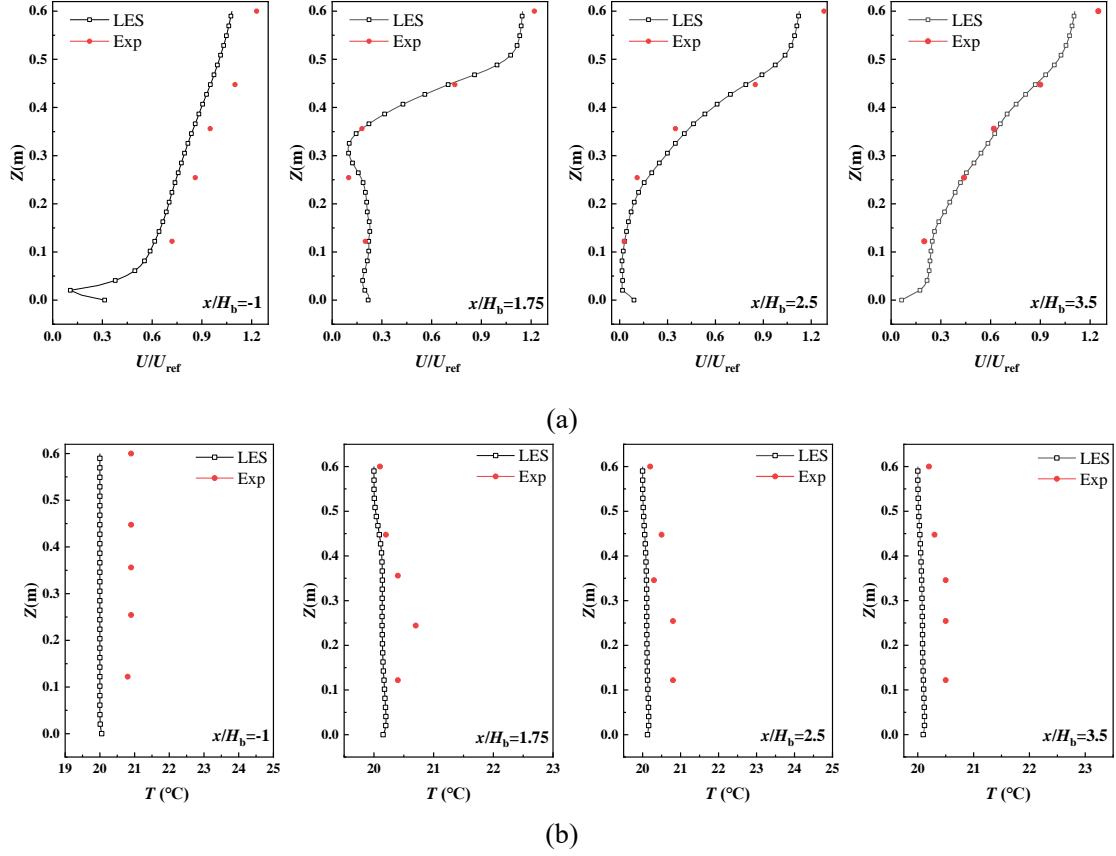

**Fig. 5.** Comparison of LES results and wind tunnel tests of (a) normalized mean velocity magnitude for case ISO-0.5 and (b) mean temperature for case WIN-2.0-45 in horizonal plane at $y/H_b = 0.5$.

## 4. Results

4.1 Non-heated walls

To provide a reference for the airflow patterns discussed in the subsequent sections, we will briefly examine the airflow patterns around an isothermal building. Fig. 6 presents the airflow patterns on the vertical central plane at $y/H_b = 0.5$ and the horizontal plane at $z/H_b = 1$. As the incoming wind approaches the windward wall of the building, the airflow disperses progressively upward, downward, and laterally. The stagnation point is observed at approximately 2/3 of the building height. Around the edges of the building, the airflow experiences significant horizontal and vertical acceleration. The separated flows reattach at the roof and leeward sides. Only the airflow above the stalled streamline can bypass the building, while a distinct curved shear layer forms on the roof surface [42]. Additionally, downward-moving airflow generates vortices near the ground surface.

For ease of explanation in the following sections, we refer to the vortex at the bottom corner of the windward side as the horseshoe vortex, the separation bubble at the top of the building as the top vortex, and the vortex on the leeward wall as the wake vortex. Three-dimensional streamlines, as depicted in Fig.6, provide a visual representation of the size and shape of these vortex structures within the flow field. The top vortex sheds downstream at approximately the height of the building. The vortices generated within the shear layer fundamentally alter the flow field near the building and the recovery of the far wake. On the windward side, near the ground surface, the horseshoe vortex opposes the incoming flow direction, resulting in a low-speed region where the two opposing flows converge near the ground. The horseshoe vortex extends to both sides of the building, creating high-speed regions at the sidewall edges where lateral vortices form. A clockwise-rotating vortex exists within the recirculation region on the leeward facade of the building. From Fig.6 (b), it can be observed that the length of the recirculation region extends approximately 0.4 m behind the leeward wall of the building, and the flow outside the recirculation region returns to a normal pattern. However, within a range of approximately 1.3 m behind the leeward wall of the building, the airflow velocity remains lower than the incoming wind speed.

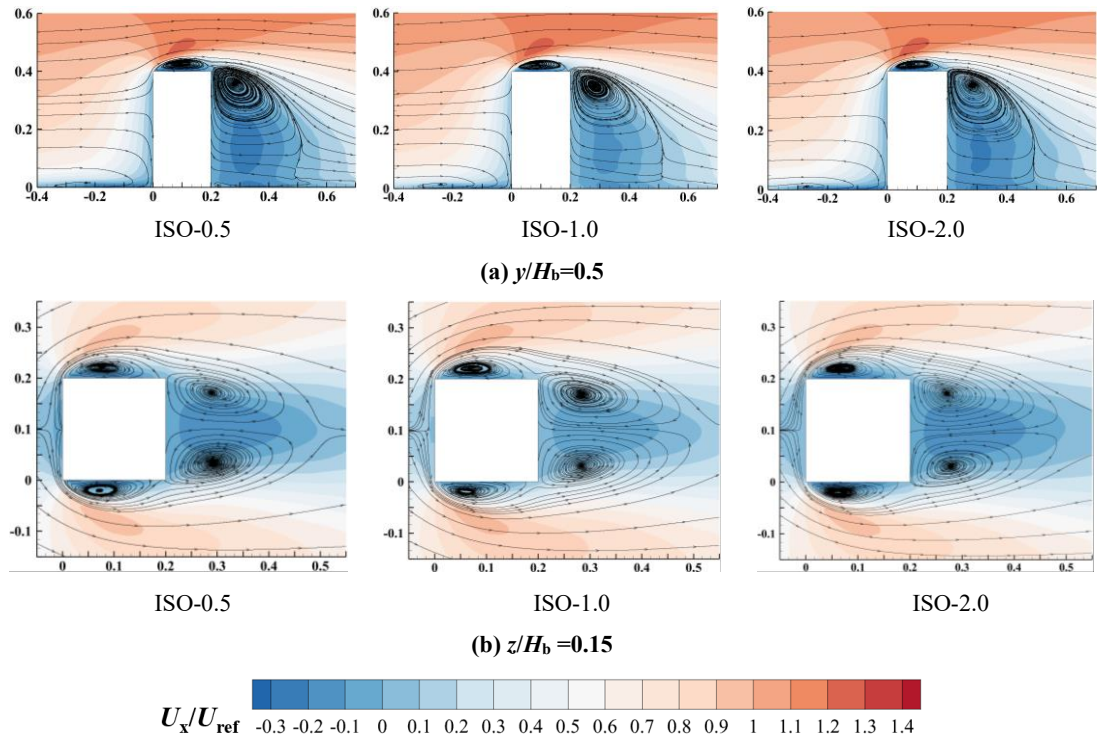

**Fig. 6.** Time-averaged airflow streamlines colored by the normalized time-averaged $U_x$ over (a) the vertical central plane at $y/H_b$=0.5, and (b) the horizontal plane at $z/H_b$ =0.15 ($Ri$=0) (the dimensions are in [m]).

Under different incoming flow velocities, the positions of the vortex cores for the top vortex and wake vortex exhibit negligible displacement (Table 3). However, the incoming wind speed has a significant impact on the length of the recirculation region on the roof, while its effect on the length of the recirculation region behind the building is relatively minor. With an increase in the incoming wind speed, the velocity gradient within the shear layer on the roof intensifies. This leads to the forward movement of the reattachment point on the roof and a reduction in the length of the recirculation region.

4.2 Leeward wall heating

Compared to isothermal conditions, heating the leeward wall induces a vertical upward motion, which affects the low-speed air within the recirculation region, as depicted in Fig. 7. This air is drawn into the heated surface and enters the thermal plume, causing a change in the flow direction near the wake, primarily moving upward towards the leeward wall. Particularly, under the $Ri$=4.00 condition, the thermal plume near the leeward wall continuously rises, leading to a small cavity behind the roof edge with a counterclockwise-rotating corner vortex. Additionally, the top vortex shows an inclined upward trend.

Furthermore, turbulent kinetic energy ($k$) serves as an indicator of the magnitude of vertical motion. Under the $Ri$=2.94 and 4.00 conditions, the magnitude of vertical velocity within the recirculation region increases significantly. As shown in Fig. 18, heating also results in a significant widening of the free shear layer's thickness due to diffusion, leading to a certain increase in $k$. The maximum turbulent kinetic energy ($k_{max}$) is observed at the trailing edge of the building roof and follows the vortex transport in a clockwise direction. However, the higher $k$ at the center of the wake suggests that the disturbance caused by vortex shedding decays from the center to the downstream.

Under $Ri$=4.00 conditions, within a range of approximately 0.05m (0.25$H_b$) outside the recirculation region, the airflow maintains a noticeable upward trend due to buoyancy effects, without an immediate return to its regular flow direction. The contribution of Reynolds normal stress on mean momentum transport is minimal, whereas shear stress takes on a leading role in the mean momentum transport induced by turbulent motion. The absolute value of the normalized vertical momentum flux $<U'W'>/U_{ref}^2$ on the roof is maximized at the locations with the largest mean velocity gradients and $k$. This is primarily due to turbulence generated by vortex shedding at the top of the building, resulting in high shear forces in regions with significant mean velocity gradients. Moreover, as the $Ri$ increases, the maximum absolute value of $<U'W'>/U_{ref}^2$ on the roof continues to increase, and the range of high momentum flux regions expands downstream. In conclusion, heating the leeward wall not only alters the magnitude of airflow velocity but also enhances $k$.

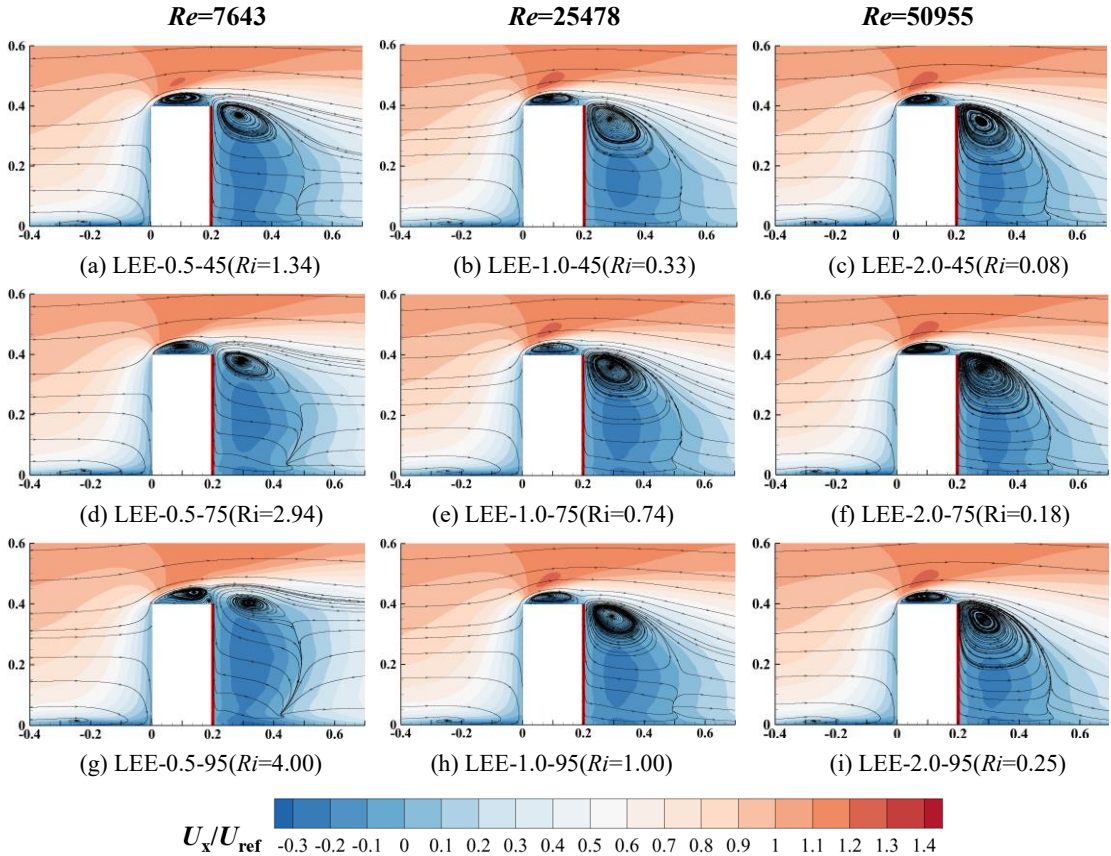

**Fig. 7.** Time-averaged airflow streamlines colored by the normalized time-averaged $U_x$ over the vertical central plane at $y/H_b$=0.5 (the heated wall is represented by a red line; the dimensions are in [m]).

The influence of heating the leeward wall on the flow characteristics is clearly observed in Fig. 8. Under both non-heated and heated leeward wall conditions, the magnitudes of the normalized mean velocities along the streamwise direction ($U_x$) and the vertical direction ($U_z$) near the roof and leeward side of the building are depicted (comparisons of $U_x$ on other measurement lines can be found in Appendix A). When the leeward wall is heated, there are no significant differences in the various conditions upstream of the building. Under the condition of $Ri$=4.00 for $x/H_b$=0.5, the wind speed in the streamwise direction decreases, while the $U_z$ significantly increases. Similarly, on the leeward wall near the wall at $x/H_b$=1.1, under low $Re$ conditions ($Re$=7643), all cases exhibit a significant increase in $U_z$ in the vertical direction. In addition, as the heating temperature increases and $U_z$ becomes larger, it means that buoyancy accelerates the flow, ultimately leading to a more noticeable upward effect of the thermal plume. This correlation implies a relationship between heating intensity and the intensified vertical motion of the airflow. Notably, under the condition of $Ri$=4.00 for $x/H_b$=1.1, at a height corresponding to the building height ($H_{ref}$), the normalized $U_z$ is 3.4 times higher than that in the non-heated condition.

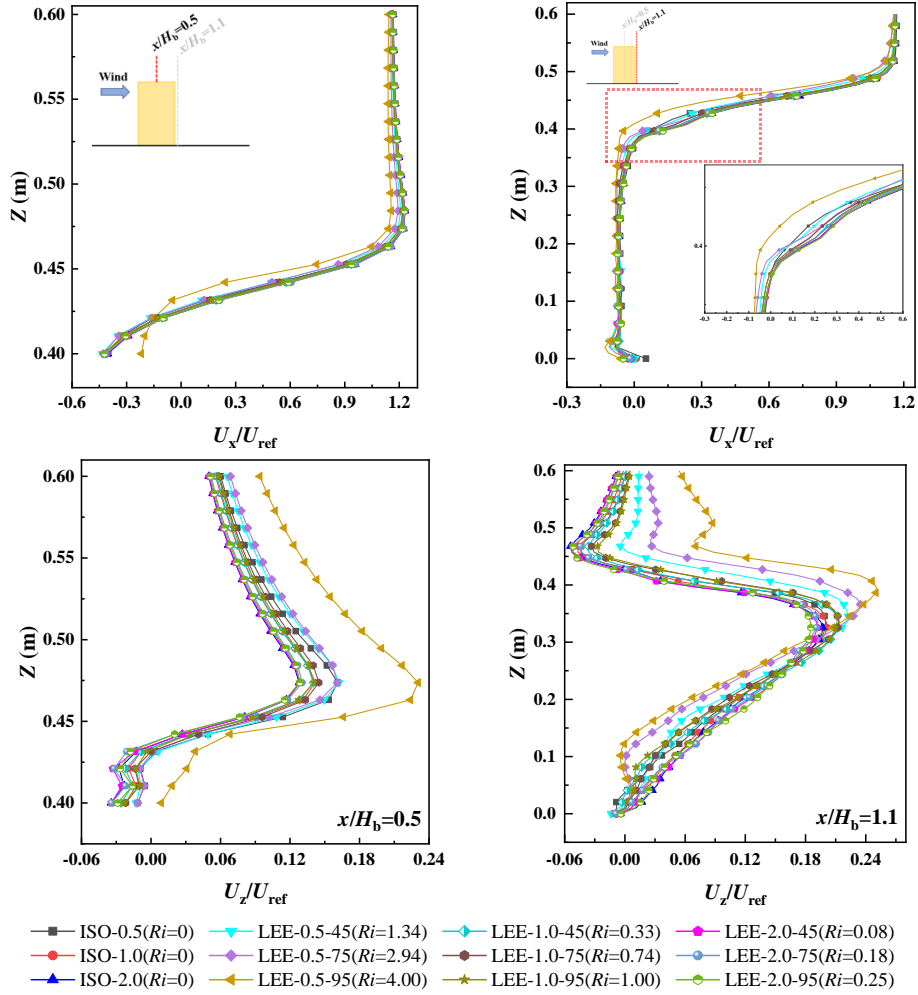

**Fig. 8.** Comparison of normalized $U_x$ and $U_z$ along vertical lines at $x/H_b$=0.5 and $x/H_b$=1.1 for different incoming wind speeds when heating the leeward wall.

Due to the influence of thermal effects on flow patterns, the lengths of the reattachment region on the roof and the recirculation region on the leeward side have undergone changes, as indicated in Table 3. When Compared to the non-heated condition, the most significant alterations occur at $Ri$=2.94 and 4.00, where the reattachment region lengths are reduced by 6% and 10.5%, respectively, and the recirculation region lengths decrease by 8.25% and 12.5%, respectively. Furthermore, under the $Ri$=4.00 condition, the vortex core of the wake is notably elevated and moves away from the leeward wall. The vortex core position of the top vortex slightly shifts towards the trailing edge of the roof, without a noticeable upward trend. Additionally, under high *Re* number conditions (*Re*=50955), mechanical flow dominates, while thermal effects have negligible influence. Therefore, the flow patterns are essentially the same as those observed under isothermal conditions. It is important to note that the flow pattern of the horseshoe vortex on the windward side remains largely unchanged across all conditions, and thus, it is not extensively discussed in this paper.

**Table 3**
Comparison of top and wake vortex locations, reattachment and recirculation region lengths for isothermal walls and heated leeward wall.

| Case | $X_1/H_b$ | $X_2/H_b$ | Vortex core 1 (m) | Vortex core 2 (m) |
|---|---|---|---|---|
| ISO-0.5 | 1.000 | 2.000 | (0.10,0.43) | (0.28,0.36) |
| ISO-1.0 | 0.945 | 2.000 | (0.09,0.43) | (0.28,0.35) |
| ISO-2.0 | 0.920 | 1.945 | (0.07,0.42) | (0.28,0.35) |
| LEE-0.5-45 | 0.955 | 1.930 | (0.10,0.43) | (0.29,0.37) |
| LEE-0.5-75 | 0.935 | 1.835 | (0.10,0.43) | (0.29,0.38) |
| LEE-0.5-95 | 0.895 | 1.750 | (0.14,0.44) | (0.32,0.40) |
| LEE-1.0-45 | 0.935 | 2.045 | (0.08,0.42) | (0.28,0.36) |
| LEE-1.0-75 | 0.955 | 1.985 | (0.09,0.43) | (0.29,0.36) |
| LEE-1.0-95 | 0.930 | 1.990 | (0.09,0.43) | (0.29,0.36) |
| LEE-2.0-45 | 0.924 | 1.940 | (0.08,0.42) | (0.28,0.35) |
| LEE-2.0-75 | 0.930 | 1.990 | (0.08,0.42) | (0.28,0.36) |
| LEE-2.0-95 | 0.920 | 1.895 | (0.08,0.42) | (0.28,0.36) |

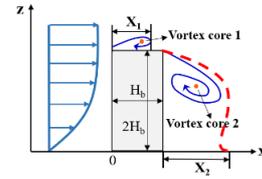

Note: $X_1$ refers to the length of reattachment region on the roof ($X_1$ represents the distance from the leading edge of the roof to the location where the near-wall airflow initiates its flow in the negative *x*-direction.); $X_2$ refers to the length of recirculation region on the leeward side; Vortex core 1 refers to the location of the top vortex core ($X_2$ indicates the distance from the windward side to the location where near-ground airflow initiates its flow in the negative *x*-direction.); Vortex core 2 refers to the location of the wake vortex core. The vortex core position refers to the plane where $y/H_b$=0.5.

When the leeward side is heated, the interaction between the incoming flow at ambient temperature on the roof and the warmer air on the leeward side causes the warmer air to be washed downstream. This phenomenon highlights the influence of heating on the transport of heat downstream. Consequently, an elongated region with higher temperatures forms at the trailing edge of the roof. This elongated hot spot is clearly visible at the trailing edge of the roof under *Ri*=2.94 and 4.00 conditions (Fig. 9 d and g). However, this phenomenon is less distinct when *Ri*<0.33.

A link exists between heat and momentum transfer: areas with increased turbulence levels promote heat transfer. Concurrently, the temperatures near the leeward wall decrease rapidly. For example, under the *Ri*=4.00 condition, the temperature drops to 26.3% of the heating temperature roughly 0.01m (0.05$H_b$) downstream of the leeward wall. Under the *Ri*=2.94 condition, the temperature decreases to 34.7% of the heating temperature at the same distance from the leeward wall. Moreover, it is noteworthy that most of the heat is not carried into the wake through the recirculation region; instead, it is vertically carried away by the thermal plume. As the thermal plume expands and rises, the influence of the heated leeward wall on the flow patterns intensifies progressively along the height of the building.

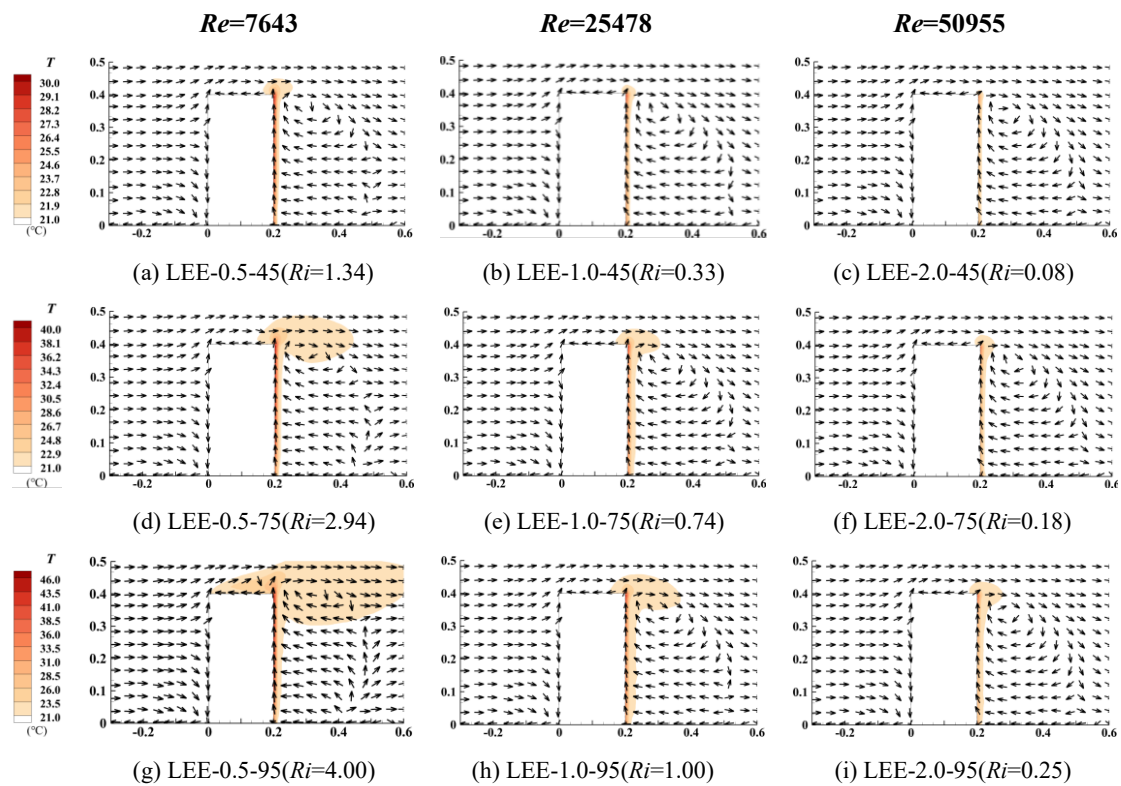

**Fig. 9.** Time-averaged velocity vectors and temperature over the vertical central plane at $y/H_b$=0.5 (the dimensions are in [m]).

By examining the horizontal temperature distribution at pedestrian height, we can gain insights into the urban heat island effect that is prevalent within cities [43]. In this study, a specific height of 1.5 meters was chosen to investigate the temperature distribution at pedestrian height level [44, 45].The introduction of heat on the leeward side demonstrates negligible influence over the temperature distribution at this height. Across all simulated cases, a marginal elevation in airflow temperature is observed adjacent to the leeward wall. Due to the lower airflow velocity on the leeward side, the thermal airflows also do not exhibit a downstream flow. It is essential to highlight that, owing to limitations in the article's length, this chapter excludes the illustration of temperature contours at pedestrian height in cases where heat is applied to the leeward side.

### 4.3 Windward wall heating

In contrast to heating the leeward wall, the heating of the windward wall exhibits distinct characteristics. As depicted in Fig. 10, under conditions of low $Re$ ($Re$=7643), the thermal plume displays two separate trajectories. A segment of the plume is directed towards the leeward side from the lateral direction, affected by the incoming airflow. Simultaneously, another segment of the plume enters the roof shear layer. In the cases of $Ri$=2.94 and 4.00, the thermal plume enters the wake region from the lateral side, and low-speed air is drawn into the thermal plume. It can be observed that the flow direction of near wake is upward towards the leeward wall (Fig. 10 d and g). The velocity and temperature field distributions at high $Re$ ($Re$=50955) can be found in Appendix B.

When heating the windward side, there is a certain degree of increase in both the thickness of the free shear layer and the magnitude of $k$. However, this increase is not as pronounced as observed when heating the leeward wall or all walls. As shown in Fig. 18(a), the presence of $k_{max}$ is observed in the middle of the roof, and it is transported along the vortex in a clockwise direction. The magnitude of *k* on the roof surface is similar to the non-heated condition. Specifically, a region with higher *k* is observed near the leeward side close to the ground, which is caused by the enhanced airflow disturbance resulting from the collision of the thermal plume after flowing into the leeward side from both sides of the building.

The absolute value of the normalized momentum flux $<U'W'>/U_{ref}^2$ on the roof surface peaks and it does not differ significantly from the non-heated condition, as shown in Fig. 18(b). In the case of *Ri*=4.00, the high-momentum flux region extends to 0.8m ($4H_b$) downstream from the leeward side. Heating the windward wall has little effect on the flow characteristics in the upstream region of the building. It mainly transfers the heated air to the downstream of the building through the incoming flow, thereby altering the flow characteristics in the wake region.

When the windward wall is heated, the low-speed incoming flow carries the hot airflow towards the top of the building and the leeward side. Under the *Ri*=2.94 condition, the air near the building's roof and the leeward side close to the ground is heated. As depicted in Fig.10(b), the temperature of the airflow near the leading edge of the roof can reach 25 °C, and within the recirculation region, a plume of warm air has a temperature of around 22 °C. In the case of *Ri*=4.00, the thermal plume near the windward wall, in contact with the surface, is strong enough to counteract the downward inertial forces and rises entirely from bottom to top. The temperature of the airflow on the building's roof and throughout the recirculation region increases to some extent. The temperature near the leading edge of the roof is approximately 27 °C, and on the leeward side near the ground, there is a slight temperature increase compared to the surrounding airflow, with a temperature of approximately 24 °C. The airflow temperature at other positions within the recirculation region is approximately 22 °C. The temperature near the windward wall diminishes rapidly. Under the *Ri*=2.94 condition, at a distance of about 0.01m ($0.05H_b$) front the windward wall, the temperature drops to 30.5% of the heating temperature. Under the *Ri*=4.00 condition, the temperature drops to 28.3% of the heating temperature at the same distance from the windward wall. Most of the heat is entrained into the near wake through the recirculation region, resulting in temperature accumulation on the leeward side.

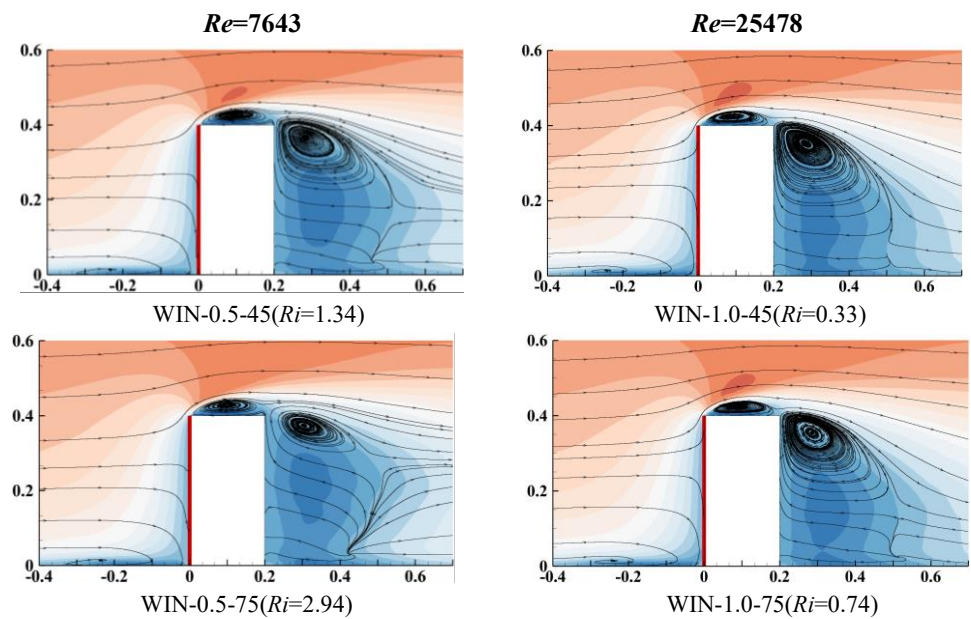

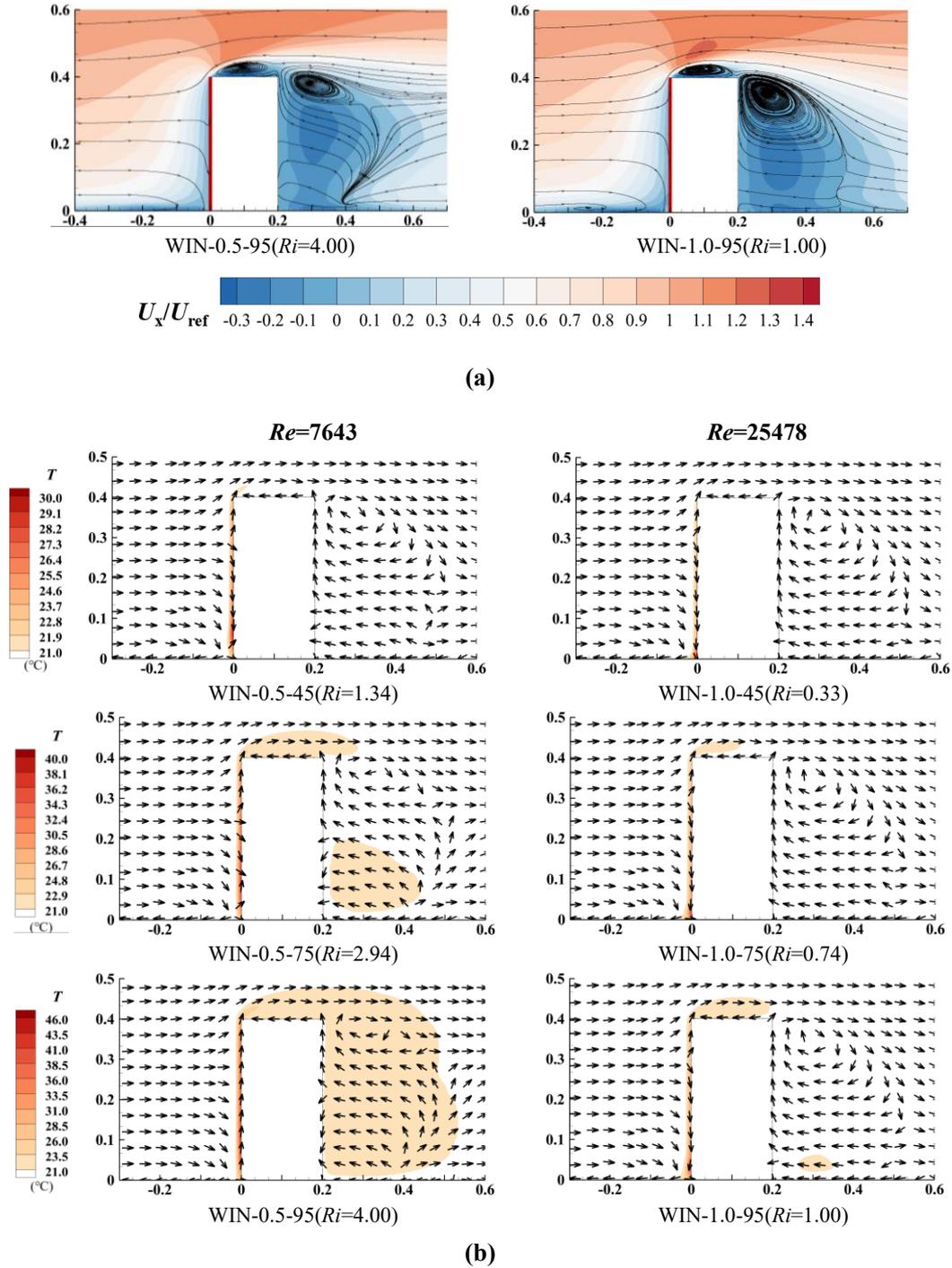

**Fig. 10.** (a)Time-averaged airflow streamlines colored by the normalized time-averaged $U_x$ and (b) time-averaged velocity vectors and temperature over the vertical central plane at $y/H_b=0.5$ (the heated wall is represented by a red line; the dimensions are in [m]).

Under both isothermal and heated windward wall conditions, the normalized mean velocities in the streamwise and spanwise directions near the roof and leeward wall of the building are depicted in Fig.11 (comparisons of $U_x$ on other measurement lines can be found in Appendix A). When the windward wall is heated, there is little variation among the cases in the upstream region of the building. For all heated windward wall conditions, the magnitude of $U_x$ remains relatively constant at $x/H_b=0.5$ and $x/H_b=1.1$. However, at low $Re$ ($Re=7643$), a significant increase in the $U_z$ is observed at $x/H_b=0.5$. Under the $Ri=4.00$ condition for $x/H_b=0.5$, the maximum $U_z$ occurs at a height of 0.47m

(2.35 $H_b$) and is 1.15 times higher than the non-heated condition. Comparatively, under the $Ri$=4.00 condition for $x/H_b$=1.1, downward airflows are observed in the height range of 0.07-0.17m (0.35-0.85 $H_b$), and under the $Ri$=2.94 condition for $x/H_b$=1.1, downward airflows are observed in the range of 0.08-0.14m (0.4-0.7 $H_b$). The magnitude of the downward airflow is larger under the $Ri$=4.00 condition. This might be because the thermal plume, which enters the wake region from the lateral side, is at a certain distance from the leeward wall. Initially, the thermal plume moves upward at an angle towards the leeward wall and then descends along with the nearby cold air as they interact. The locally heated thermal plume in the near wake is not strong enough to induce widespread upward motion. No downward airflow is observed in the vertical direction for the other conditions. It is noteworthy that above the building height, there is a significant increase in the vertical velocity of the airflow due to the impact of the thermal plume on the roof. Furthermore, as the heating temperature rises, the magnitude of $U_z$ also increases.

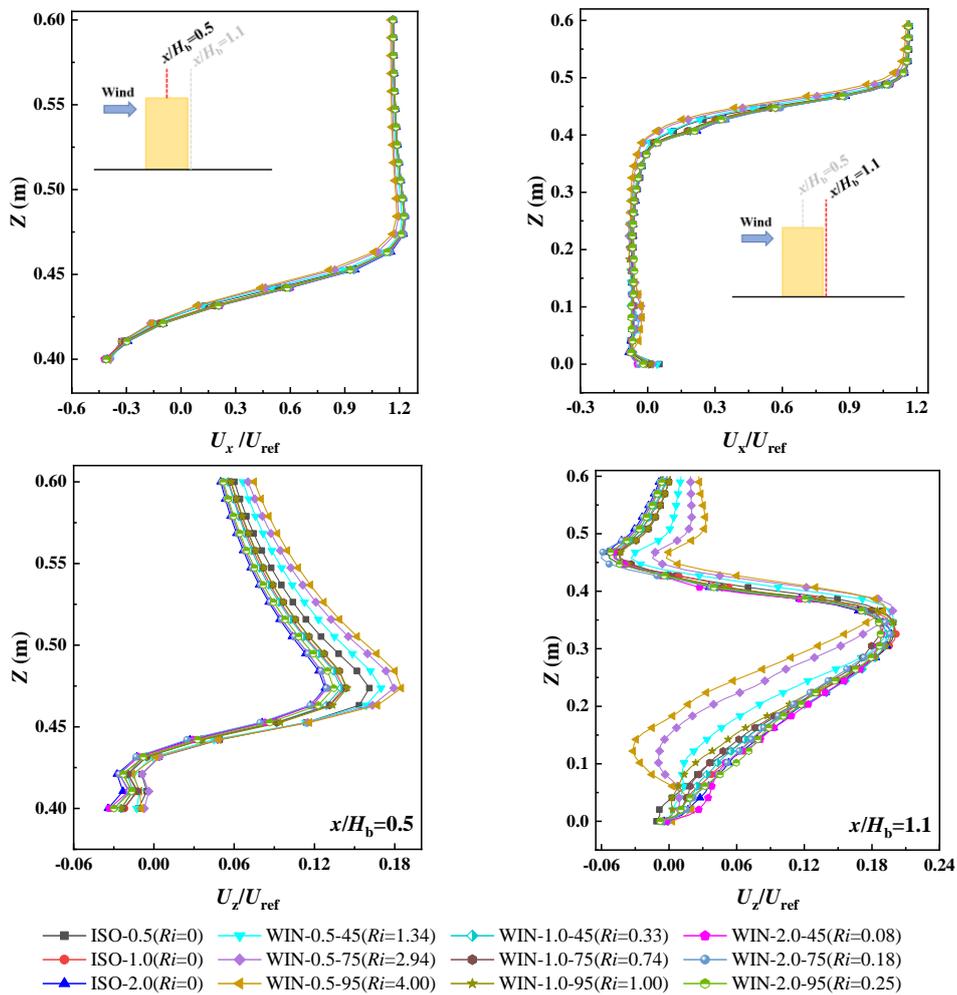

**Fig. 11.** Comparison of normalized $U_x$ and $U_z$ along the vertical lines at $x/H_b$=0.5 and $x/H_b$=1.1 for different incoming wind speeds when heating the windward wall.

At low $Re$, the length of the reattachment region on the roof shows a slight increase, while the length of the recirculation region on the leeward side experiences a significant decrease, as indicated in Table 4. In comparison to the non-heated condition, the $Ri$=2.94 and 4.00 conditions result in an increase of 6.5% and 8.5%, respectively, in the reattachment region length, while the recirculation region length decreases by 13.5% and 19.5%, respectively. The length of the recirculation region is greater than that observed in the heated leeward wall condition at the same $Ri$. In contrast to the non-heated condition, the vortex core of the wake initially moves away from the leeward wall under the $Ri$=2.94 condition but then moves back towards the leeward wall as the heating temperature of the windward wall increases to 95 °C ($Ri$=4.00). This behavior could be attributed to an increased influx of thermal plume from the building's lateral side, which promotes the movement of the near wake towards the leeward wall.

**Table 4**
Comparison of top and wake vortex locations, reattachment and recirculation region lengths when heating the windward wall.

| Case | $X_1/H_b$ | $X_2/H_b$ | Vortex core 1 (m) | Vortex core 2 (m) |
|---|---|---|---|---|
| WIN-0.5-45 | 1.020 | 1.930 | (0.09,0.43) | (0.28,0.37) |
| WIN-0.5-75 | 1.065 | 1.735 | (0.10,0.43) | (0.30,0.37) |
| WIN-0.5-95 | 1.085 | 1.610 | (0.10,0.43) | (0.29,0.38) |
| WIN-1.0-45 | 0.950 | 2.015 | (0.09,0.43) | (0.29,0.35) |
| WIN-1.0-75 | 0.949 | 1.990 | (0.09,0.43) | (0.29,0.35) |
| WIN-1.0-95 | 0.945 | 1.970 | (0.08,0.43) | (0.29,0.35) |
| WIN-2.0-45 | 0.925 | 1.965 | (0.08,0.42) | (0.28,0.35) |
| WIN-2.0-75 | 0.925 | 1.950 | (0.08,0.42) | (0.28,0.35) |
| WIN-2.0-95 | 0.925 | 1.940 | (0.08,0.42) | (0.28,0.35) |

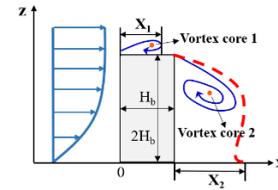

**Note:** $X_1$ refers to the length of reattachment region on the roof ($X_1$ represents the distance from the leading edge of the roof to the location where the near-wall airflow initiates its flow in the negative x-direction.); $X_2$ refers to the length of recirculation region on the leeward side; Vortex core 1 refers to the location of the top vortex core ($X_2$ indicates the distance from the windward side to the location where near-ground airflow initiates its flow in the negative x-direction.); Vortex core 2 refers to the location of the wake vortex core. The vortex core position refers to the plane where $y/H_b$=0.5.

Based on the results in Fig. 10, when the windward wall is heated, the low-speed incoming airflows at ambient temperature carry the high-temperature airflows from the windward side to the upper and leeward side of the building. Below the point of stagnation, the airflow moves along both sides towards the leeward side. Compared to the non-heated condition, in the cases with an incoming wind speed of 0.5m/s, as the wall heating temperature increases, the extent of the low wind-speed zone near the wake region of the isolated building gradually diminishes. As depicted in Fig. 13, at incoming wind speeds of 0.5 m/s and 1 m/s, an elevated temperature on the windward side leads to a gradual movement of heated airflow along both sides of the building towards the leeward side. Consequently, a subtle heating effect emerges within the airflow of the nearby wake region. Nevertheless, as the incoming wind speed increases to 2 m/s, even though the heating temperature on the windward wall is heightened and heated airflows towards both sides of the building, the impact of the heated airflow on the temperature of the airflow in the nearby wake region remains minimal. This limited effect can be attributed to the higher wind speed, which causes a rapid decrease in the temperature of the heated airflow, consequently constraining its effective reach.

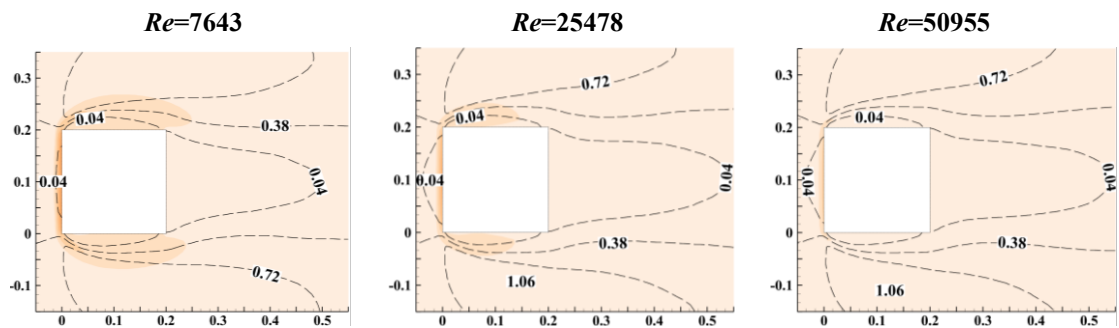

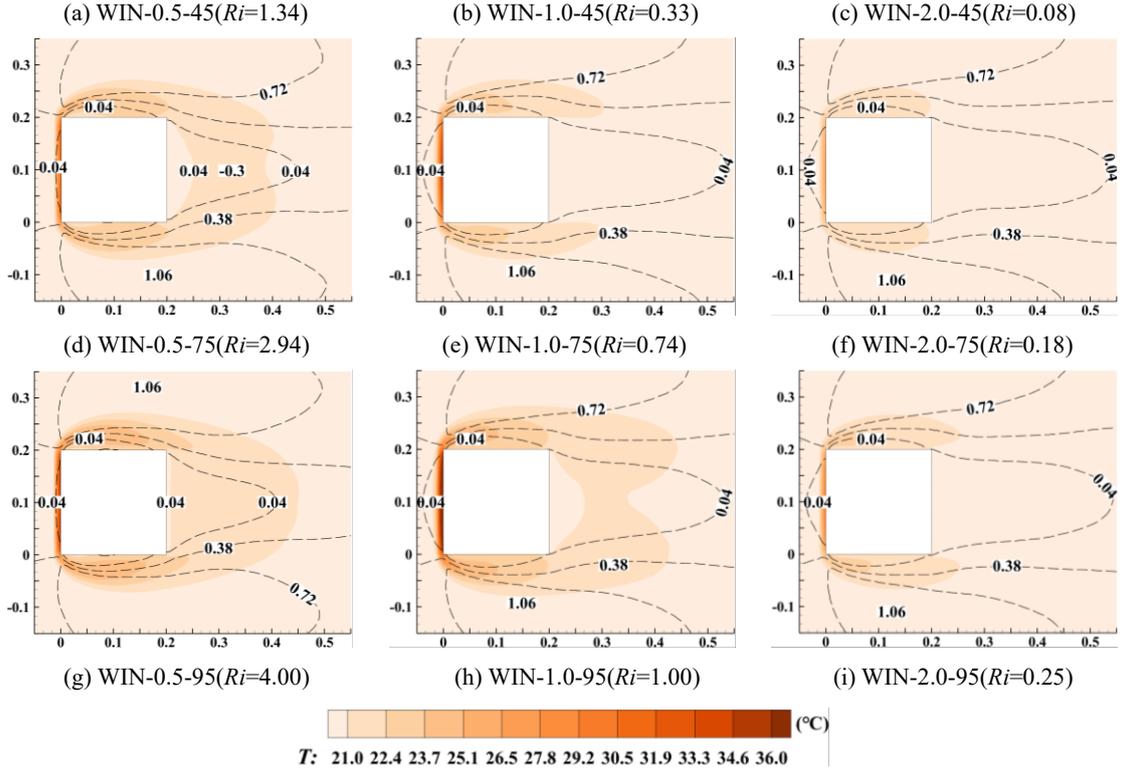

**Fig. 12.** Time-averaged temperature over the horizontal plane at $z/H_b$=0.15 (the long-dashed lines represent the contour of the normalized time-averaged $U_x$; the dimensions are in [m]).

### 4.4 All-walls heating

When all walls are heated (including side walls), the flow structures on the vertical plane undergo changes due to the increasing buoyancy. These changes are particularly notable at low *Re*. In the case of *Ri*=1.34, the wake vortex is disrupted by the buoyancy flow moving towards the windward wall, while the top vortex remains intact. Despite that, the deformation of the flow field caused by mixed convection has already occurred in the near wake, and the recovery of the flow field in the far wake requires the formation of new attached starting points, which are clearly visible in Fig. 13. For *Ri*=2.94 and 4.00, both the wake and top vortex structures are disrupted, and the vortex structures cease to exist. The stronger the vortices, the higher the inertial forces, necessitating a stronger buoyancy flow to dissipate the vortices.

As shown in Fig. 18, regions with high *k* appear near the trailing edge of the roof and in the near-ground region of the reattachment region. The shear layer that separates from the leading edge of the roof reattaches near the trailing edge, and the shedding of vortices contributes to an increase in *k*. The augmentation of turbulence in the near-ground region of the reattachment region is caused by the collision of hot airflows entering from both lateral sides on the leeward side. Due to the absence of the top vortex structure, the $k_{max}$ emerges in the near-ground of the reattachment region, and the *k* at the trailing edge of the roof is comparable to the case without heating. The maximum absolute value of $<U'W'>/U_{ref}^2$ still persists at the trailing edge of the roof and within the free shear layer, but its position does not coincide with that of $k_{max}$. As the distance from the building increases, the correlation between the streamwise and spanwise velocities diminishes, resulting in the normalized Reynolds shear stress $<U'W'>/U_{ref}^2$ approaching zero. Simultaneously, $<U'W'>/U_{ref}^2$ within the recirculation region amplifies with increasing temperature, and the region of high-momentum flux inclines towards the free shear layer and develops downstream. Under other conditions, although the positions and velocities of the vortices exhibit some variations, the vortex structures continue to exist. Comparatively, heating all walls yields more pronounced alterations in the flow field structure than solely heating the windward or leeward wall. Particularly at high $Ri$, the reshaping of the flow field structure on the vertical plane is achieved.

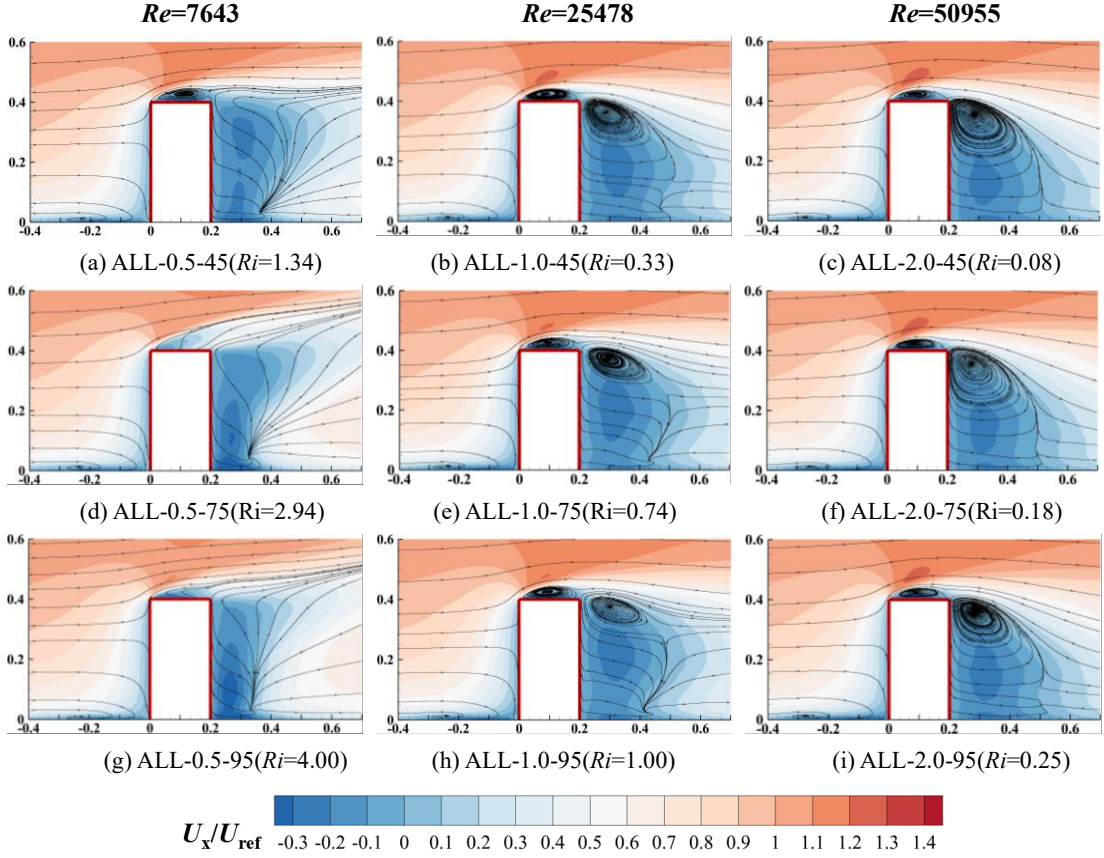

**Fig. 13.** Time-averaged airflow streamlines colored by the normalized time-averaged $U_x$ over the vertical central plane at $y/H_b=0.5$ (the heated wall is represented by a red line; the dimensions are in [m]).

When all walls are heated, there are only slight variations among different conditions upstream of the building, while more significant differences are observed on the roof surface and the leeward side (comparisons of $U_x$ on other measurement lines can be found in Appendix A). Along the $x/H_b=0.5$ measurement line, within a distance of 0.02m ($0.1H_b$) from the roof, the airflow moves in the positive x-direction only in the case of $Ri=2.94$. Due to the influence of the thermal plume from the roof, the velocity in this case is higher than in the other cases. Additionally, the temperature decay is slower within the 0.02m range from the roof. The magnitude of $U_z$ reaches its peak at a position 0.02m from the roof in the $Ri=2.94$, and then gradually decreases, approaching the values of the other conditions. Beyond the 0.02m range from the roof, the $U_x$ is lower in the $Ri=2.94$ case compared to the other cases. Along the $x/H_b=1.1$ measurement line, within the height range of 0.45-0.5m ($2.25$-$2.5H_b$), there is a sharp decrease in $U_x$ and a significant increase in the corresponding $U_z$ in the $Ri=2.94$ case. This phenomenon is clearly observed when examining Fig. 15 and Fig. 16. In the cases of $Ri=2.94$ and 4.00, there have been notable changes in the flow pattern of $U_z$, exhibiting significant differences compared to the other conditions.

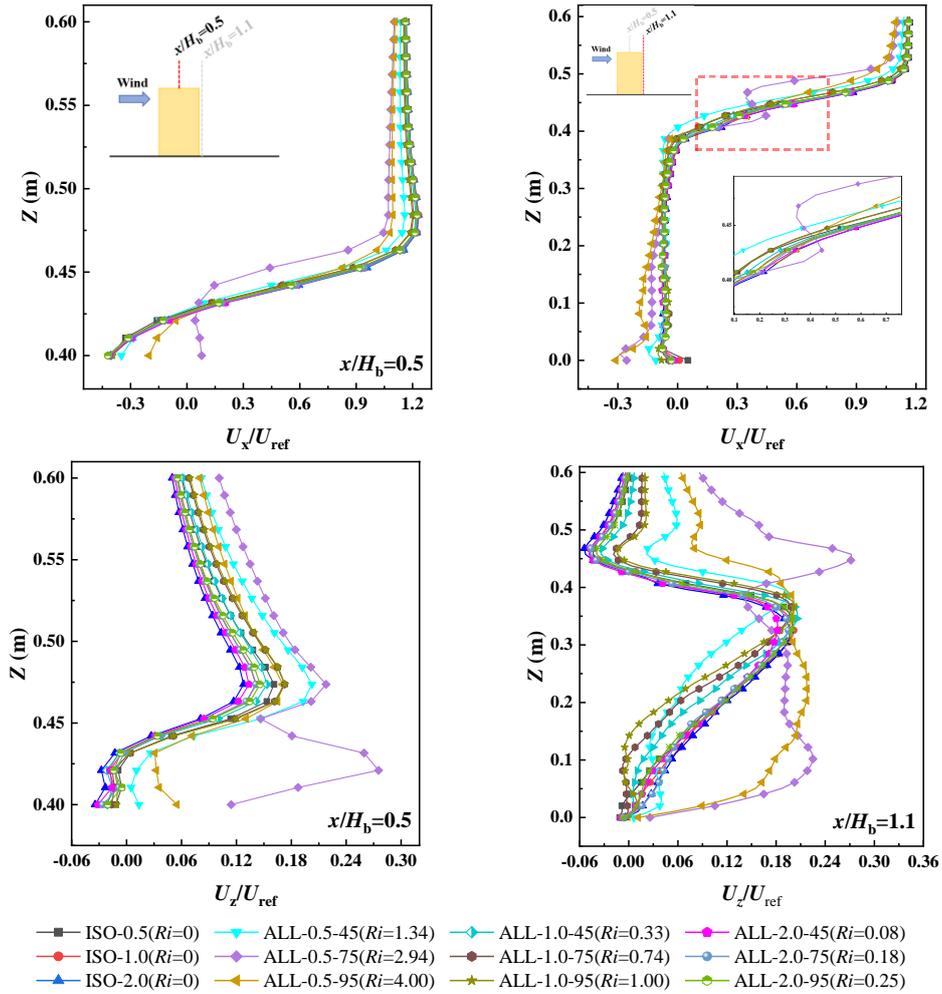

**Fig. 14.** Comparison of normalized $U_x$ and $U_z$ along the vertical lines at $x/H_b$=0.5 and $x/H_b$=1.1 for different incoming wind speeds when heating all walls.

The heating of all walls under the influence of strong buoyancy flows induces changes in the structure of the flow field. As indicated in Table 5, in comparison to the non-heating case, the $Ri$=2.94 and 4.0 conditions do not exhibit a roof reattachment region (no reattachment point is defined), and the recirculation region is reduced by approximately 48.3% and 46.8%, respectively, which is about half of the non-heating case. However, in the $Ri$=1.34 condition, the roof reattachment region slightly increases while the recirculation region decreases by 38.3%. For the $Re$=25478 and 50955 cases, the roof reattachment region experiences a certain degree of enlargement, and there is a tendency for the recirculation region on the leeward side to contract. The position of the top vortex core remains almost unchanged, except for the $Ri$=1.00 case where the position of the wake vortex core noticeably rises and moves away from the leeward wall. In the remaining conditions, the position of the wake vortex core remains almost the same.

**Table 5**
Comparison of top and wake vortex locations, reattachment and recirculation region lengths when heating the windward wall.

| Case | $X_1/H_b$ | $X_2/H_b$ | Vortex core 1 (m) | Vortex core 2 (m) |
|---|---|---|---|---|
| ALL-0.5-45 | 1.056 | 1.235 | (0.11,0.43) | -- |
| ALL-0.5-75 | -- | 1.035 | -- | -- |
| ALL-0.5-95 | -- | 1.065 | -- | -- |
| ALL-1.0-45 | 0.970 | 1.970 | (0.09,0.43) | (0.29,0.36) |
| ALL-1.0-75 | 1.015 | 1.750 | (0.09,0.43) | (0.29,0.37) |
| ALL-1.0-95 | 1.016 | 1.650 | (0.09,0.43) | (0.30,0.38) |
| ALL-2.0-45 | 0.925 | 1.975 | (0.09,0.42) | (0.28,0.35) |
| ALL-2.0-75 | 0.935 | 1.960 | (0.09,0.43) | (0.28,0.36) |
| ALL-2.0-95 | 0.950 | 1.965 | (0.09,0.43) | (0.29,0.36) |

**Note:** $X_1$ refers to the length of reattachment region on the roof ($X_1$ represents the distance from the leading edge of the roof to the location where the near-wall airflow initiates its flow in the negative $x$-direction.); $X_2$ refers to the length of recirculation region on the leeward side; Vortex core 1 refers to the location of the top vortex core ($X_2$ indicates the distance from the windward side to the location where near-ground airflow initiates its flow in the negative $x$-direction.); Vortex core 2 refers to the location of the wake vortex core. The vortex core position refers to the plane where $y/H_b=0.5$.

In the conditions where all walls are heated, at $Ri$=0.08, the airflow temperature remains unchanged within the recirculation region. However, in the other conditions, there is a noticeable increase in temperature on the roof and leeward side of the building. Specifically, in the case of $Ri$=2.94, the temperature at the leading edge of the roof is elevated. This results from a combined effect of heating the roof and the incoming wind transporting the high-temperature airflow from the windward side to the top of the isolated building. Consequently, influenced by the thermal plume, the airflow over the roof undergoes a transformation, adopting an oblique upward motion. Simultaneously, within the recirculation zone, the temperature of the airflow increases due to the synergistic impact of the heated leeward wall and the warm airflow conveyed from the windward side. This phenomenon subsequently induces a widespread upward motion of the airflows within the recirculation zone. However, the temperature rise on the leeward side is not as significant as on the roof surface. Similarly, from Fig.15(g), it can be observed that in the $Ri$=4.00 condition, there is a prevalent upward motion of the airflow near the windward wall. The temperature near the leading edge of the roof and within the recirculation region is higher, indicating a more pronounced influence of buoyancy flow. However, it is worth noting that due to the high temperature at the leading edge of the roof, a portion of the airflow near the roof surface moves towards the warmer leading edge, while another portion is entrained along with the upward flow of the thermal plume near the leeward wall, resulting in an upward movement. A distinct dividing point can be observed near the trailing edge of the roof, approximately 0.17m (0.85$H_b$) away from the leading edge of the roof. By analyzing Fig. 15 and Fig.16, it becomes evident that the airflow before the dividing point initially moves toward the leading edge of the roof and then transitions to an oblique upward motion. As the $Ri$ decreases, the temperature and range of the heated airflow on the roof and leeward side gradually diminish, and the significance of buoyancy flow decreases, giving way to mechanical flow dominating the flow characteristics.

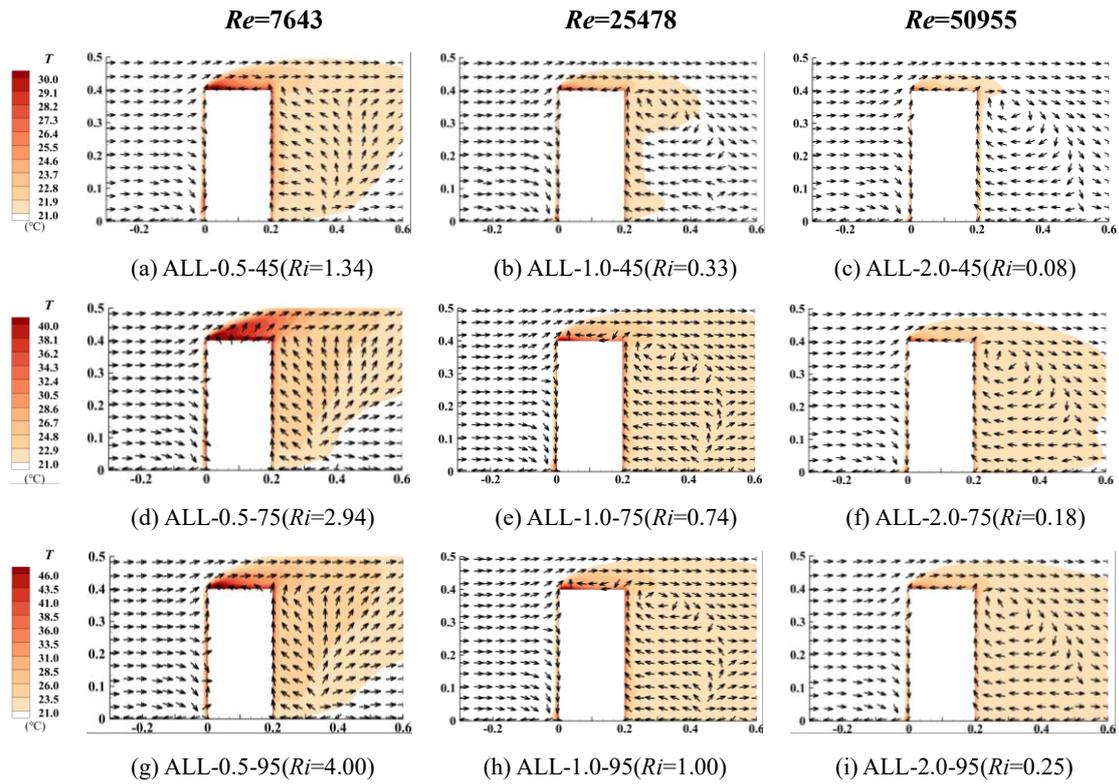

**Fig. 15.** Time-averaged velocity vectors and temperature over the vertical central plane at $y/H_b$=0.5 (the dimensions are in [m]).

In light of the combined influence of different heating temperatures and incoming wind speeds, the temperature distribution at pedestrian height is depicted in Fig.16 when all walls are heated. Upon heating all walls, a notable synergistic relationship between temperature and incoming wind speed becomes evident, resulting in a spatial overlap between hot spots and regions characterized by low wind speeds. When the wall heating temperature is relatively high, especially at the four corner positions of the isolated building, local hot spots are formed due to the lower wind speed. At an incoming wind speed of 0.5 m/s, the thermal airflow primarily converges at the four corner positions of the isolated building and within the wake region. In contrast to the non-heated cases, with an increase in the wall heating temperature, the size of the low wind-speed zone near the wake region of the isolated building notably diminishes. Nevertheless, with the escalation of incoming wind speeds to 1 m/s and 2 m/s, and coupled with an increase in wall heating temperature, the thermal airflow extends downstream of the isolated building. Of particular note, under the condition of $Ri$=1.00, the thermal airflow at the pedestrian height level has the widest range of effects, while the hot spots around the isolated building show the greatest range.

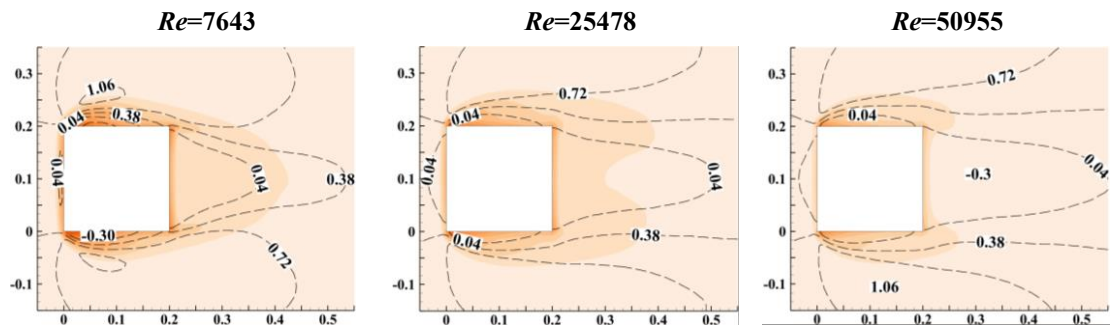

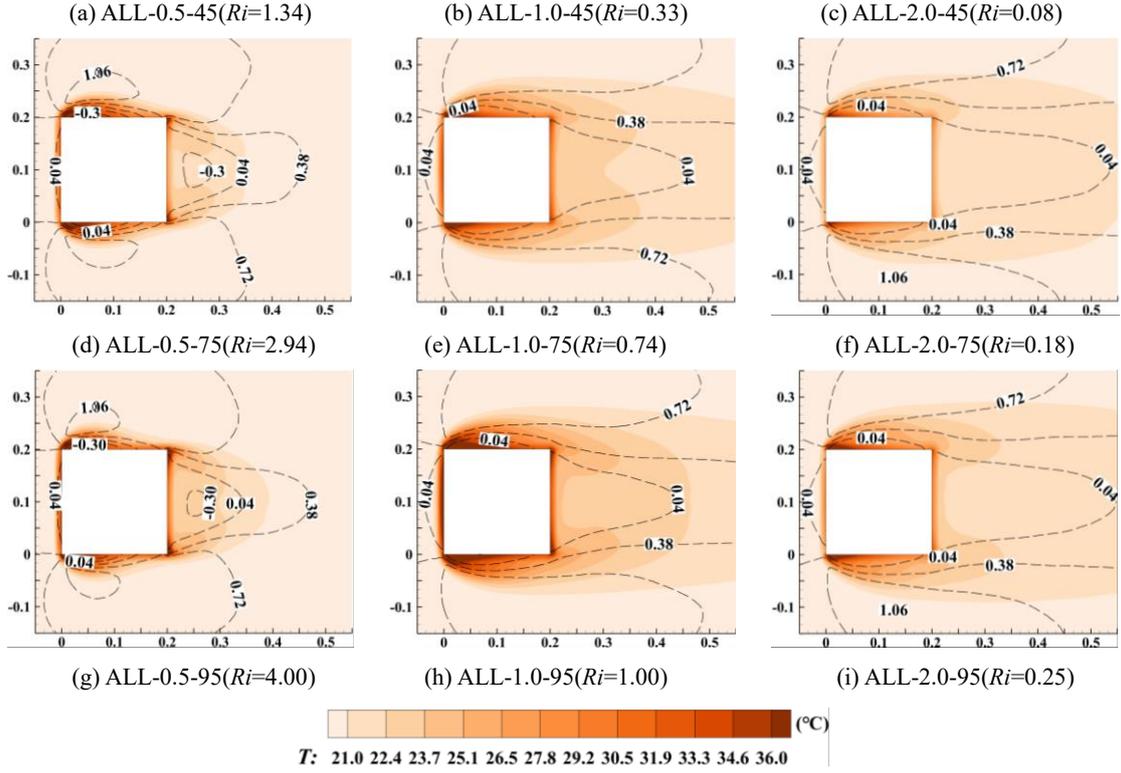

**Fig. 16.** Time-averaged temperature over the horizontal plane at $z/H_b$=0.15 (the long-dashed lines represent the contour of the normalized time-averaged $U_x$; the dimensions are in [m]).

## 5. Discussion

We explored the flow field around a building, focusing on varying thermal effects, using Large-Eddy Simulations and wind tunnels measurements where the building surfaces were heated. The relative magnitudes of the $Ri$ were controlled by setting appropriate wall heating temperatures and selecting representative speeds of the approaching flow. Conducting studies in this manner enables the discovery of general flow patterns and underlying phenomena. While this approach might not necessarily replicate the temperature distribution in real-world outdoor environments, it allows for better control over the building surface temperature and guarantees repeatability [46]. Furthermore, data from wind tunnel test that incorporate thermal effects can be used to validate numerical simulation results. The LES method is capable of accurately simulating the wind and thermal conditions surrounding the isolated building. It allows for capturing the turbulent nature of the airflow around the isolated building, providing a holistic understanding of the phenomenon.

In terms of non-heated conditions, comparing the flow field structures under three different incoming wind speeds, it is observed that the overall flow pattern does not significantly change with variations in the incoming wind speed. The most significant impact is observed in the lengths of the reattachment and recirculation regions [47]. However, it is important to note that the size of the recirculation region on the leeward side is not solely determined by the upstream flow rate [48].

Buoyant flow can alter the flow patterns around the isolated building. While moderate heating of the building surfaces has a minimal impact on the surrounding flow field, vortex circulation within the recirculation region does accelerate with increased heating temperatures. Comparing the lengths of the recirculation region under identical incoming wind speed and heating temperature reveals that all-wall heating results in the most significant reduction, potentially halving the length compared to the non-heated condition. This is followed by heating the windward wall, with the leeward wall heating producing the least reduction in the recirculation region.

When all walls are heated, increasing the temperature initially reduces the length of the recirculation region (except for ALL-0.5-75 and ALL-0.5-95), but it then enlarges subsequently, as summarized in Fig.17(a). Heating the leeward wall diminishes the length of the reattachment region on the roof, while heating either the windward wall or all walls lead to an increase in the length of the reattachment region. From Fig. 17(a)-(c), in any case of wall heating, a decrease in $Re$ significantly reduces the length of the recirculation region, especially when all walls are heated. This observation aligns with the findings of Ruck [26] in their analysis of the flow field around a heated building model in a wind tunnel. At $Re$=50955, as shown in Fig. 17(c), heating the windward, leeward, or all walls to different temperatures has minimal influence on the lengths of both recirculation and reattachment regions. As mentioned earlier, under high $Re$ conditions, forced convection dominates, and thermal effects have minimal impact. In contrast to previous studies, there was no observed temperature accumulation in the low-speed core of the wake vortex when only the leeward wall was heated [25]. With regard to vortex core locations, both the top vortex and the wake vortex tend to move upwards and downstream as buoyancy effect becomes more pronounced.

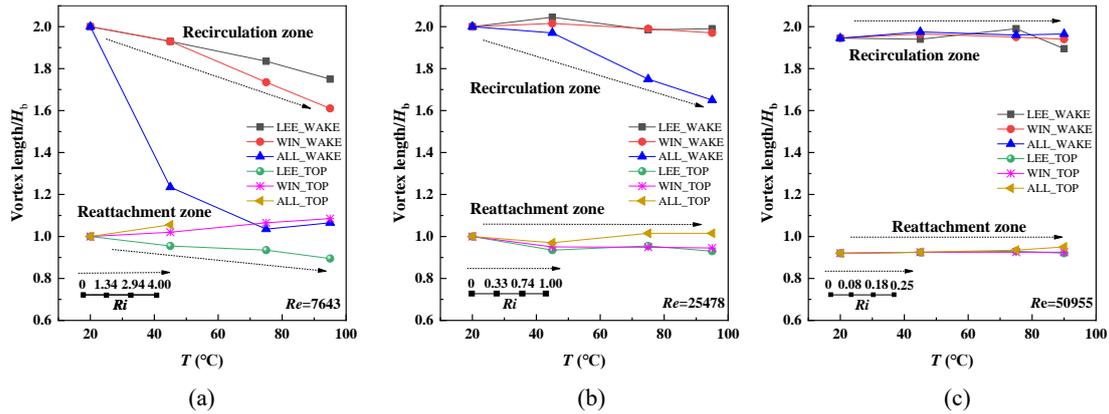

**Fig. 17.** Comparison of the lengths of the recirculation and reattachment regions when heating leeward, windward and all walls at (a) $Re$=7643 (In the ALL-TOP condition, when $Ri$ = 2.94 and 4.00, the flow structures on the roof are destroyed by buoyant flows and the reattachment region disappears), (b) $Re$=25478, (c) $Re$=50955.

For all studied conditions, heating the building walls has a minimal impact on the flow upstream of the building, and the temperature rapidly decays in the vicinity of the heated wall. Under high $Re$ ($Re$=50955), the flow characteristics of different cases are similar to the case where no heating is applied.

When the leeward wall is heated, the velocity field affects the temperature distribution within the recirculation region, and most of the heat is not entrained into the wake through the recirculation region but is carried vertically by the thermal plume. When the windward wall is heated under low $Re$, the low-speed incoming flow transports the high-temperature airflow from the windward side to the top and leeward side of the building. As for $Ri$= 4.00, the airflow temperature on the building roof and within the entire recirculation region increases to a certain extent, with most of the heat being entrained into the near wake through the recirculation region. When all walls are heated, the flow structures on the investigated vertical planes change with increasing buoyancy. At $Ri$=1.34, the wake vortex is disrupted by the buoyancy flow moving towards the leeward wall, while the top vortex remains. At $Ri$=2.94 and 4.00, both the wake and top vortex structures are disrupted, and no vortex structures are present. At the $Ri$=0.08, the airflow temperature within the recirculation region experiences only minimal increase, while in the other cases, the airflow temperatures on the building roof and leeward side increase to some extent.

In the case of low $Re$, a comparison of the highest heating temperature conditions among the four cases-non-heated, windward wall heating, leeward wall heating, and heating of all walls - reveals that the $k$ at the leading edge of the roof is relatively small for all heating conditions. There exists a region on the roof where the $k$ is larger, which gradually attenuates downstream from the position of $k_{max}$, as shown in Fig.18. For the non-heated and windward wall heating cases, $k_{max}$ is located in the middle of the roof, while in the case of leeward wall heating, $k_{max}$ appears at the trailing edge of the roof. These findings align with the conclusions drawn in the relevant studies by Murakami et al. [49, 50].

Under windward wall heating and heating of all walls, a region on the leeward side near the ground is observed where the $k$ attenuates from the inside to the outside. Moreover, in the case of heating all walls, $k_{max}$ is situated within that region. Typically, both shear forces and buoyancy contribute to the generation of turbulence. Depending solely on the mean flow in the recirculation region may not generate sufficient shear forces to account for such high turbulence levels. This could be attributed to the presence of high thermal gradients near the leeward wall, leading to increased shear forces, as well as temperature fluctuations causing buoyancy fluctuations, ultimately resulting in significant turbulence. As mentioned earlier, the high-temperature airflow on the leeward side originates partially from heating the leeward wall and mainly from the transport of heated airflow from other surfaces to the cavity behind the building under the influence of the incoming flow. A previous study also suggests a correlation between the position of $k_{max}$ and the prediction of the reattachment region [51].

By comparing Fig. 18(a) and (b), it can be observed that, the location of the absolute maximum value of $<U'W'>/U_{ref}^2$ coincides with the location of $k_{max}$. At locations where turbulent momentum flux exists in the flow field, substantial turbulence is generated through the interaction between the upper and lower layers of fluid. However, turbulence resulting from buoyancy can only cause relatively small momentum transfer. Therefore, even when all walls are heated, the vertical momentum flux near the ground on the leeward side remains lower compared to the roof level. At the building height, as the distance from the leeward wall increases, the absolute maximum value of $<U'W'>/U_{ref}^2$ gradually decreases, which can be attributed to the dissipative effects of vortex shedding [52].

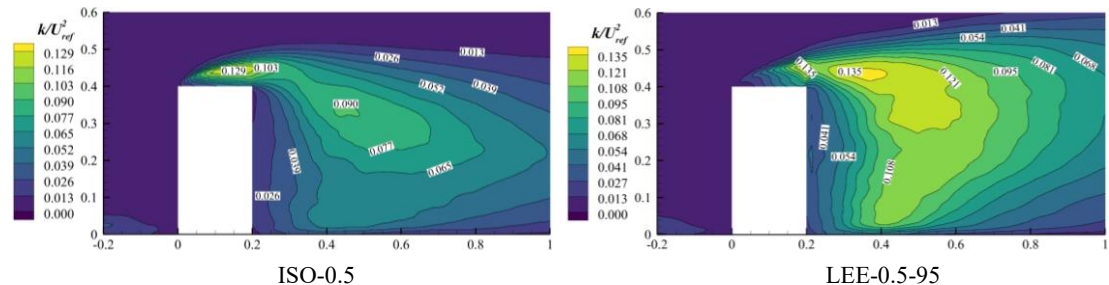

ISO-0.5　　　　　　　　　　　　LEE-0.5-95

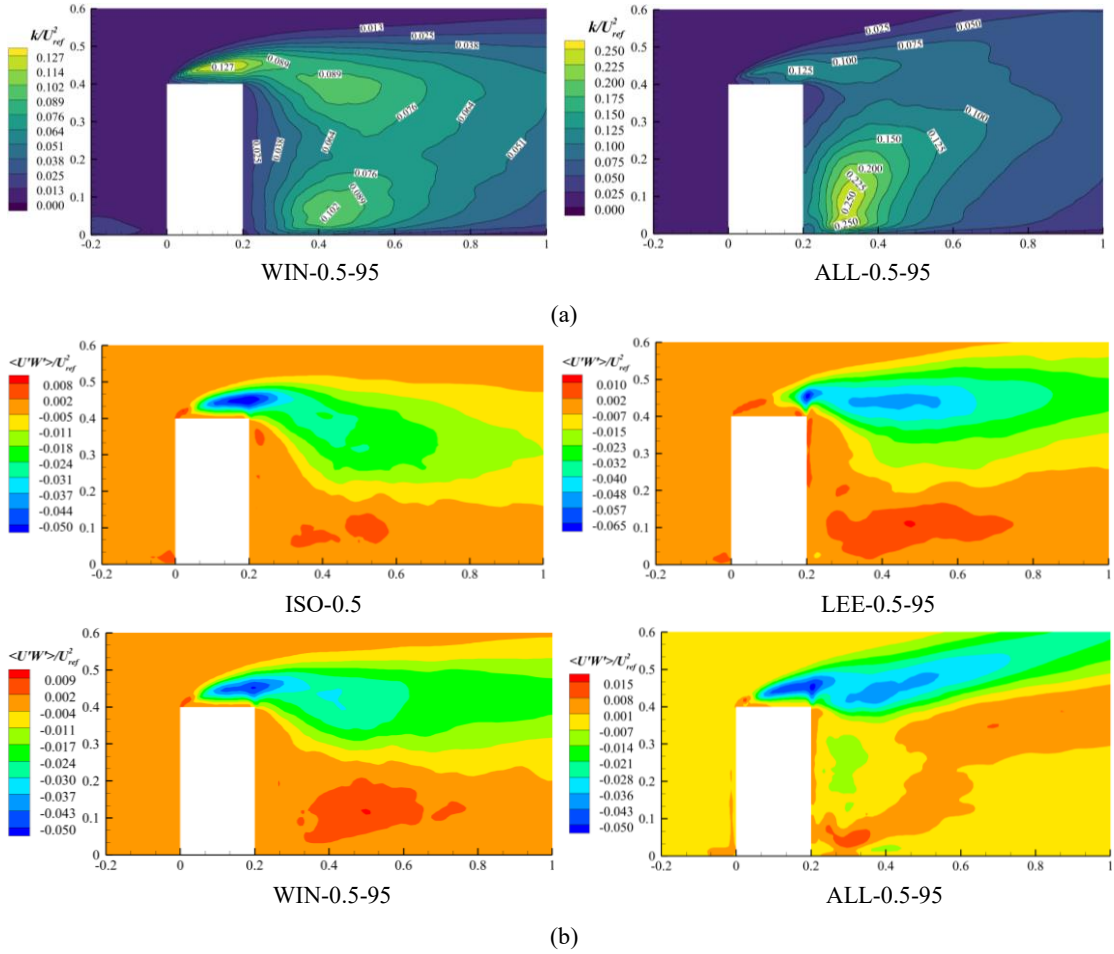

**Fig. 18.** Distribution of (a) k and (b) $<U'W'>/U_{ref}^2$ over the vertical central plane generated by four various configurations of ISO-0.5, LEE-0.5-95, WIN-0.5-95, and ALL-0.5-95 (*Re*=7643, *Ri*=4.00).

In Fig. 19, we present a visualization that depicts the distribution of vorticity magnitude across various heating conditions, which effectively identifies and locates vortices. In the absence of heating, the strength of the horseshoe vortex in front of the building is prominent. However, during heating conditions, this vortex weakens, with the phenomenon being most pronounced when all walls are heated. When heating either the leeward wall or all walls, there is a significant reduction in vortex strength near the ground on the leeward side. Simultaneously, an increase in vortex intensity is observed within the roof shear layer. This phenomenon can be attributed to the enhanced rotational motion of the airflow influenced by buoyancy.

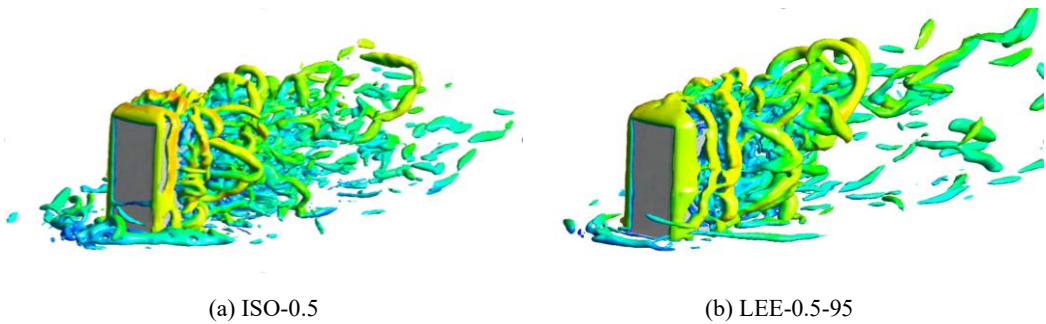

(a) ISO-0.5            (b) LEE-0.5-95

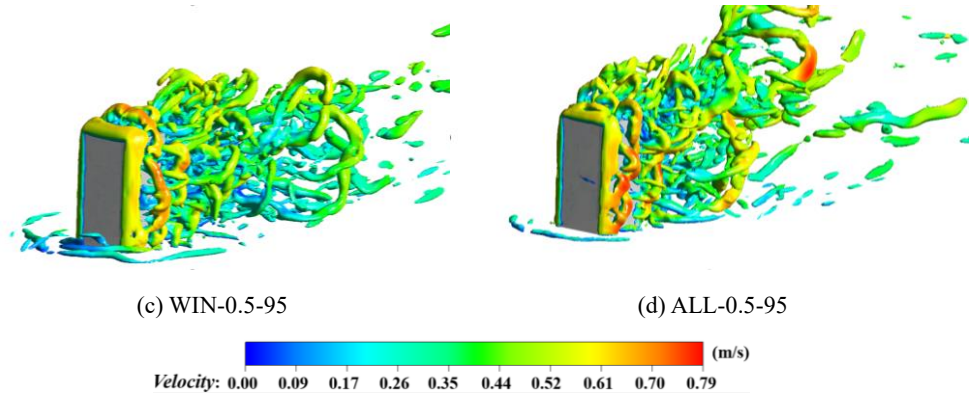

(c) WIN-0.5-95          (d) ALL-0.5-95

Velocity: 0.00 0.09 0.17 0.26 0.35 0.44 0.52 0.61 0.70 0.79 (m/s)

**Fig. 19.** Vorticity iso-surface distribution from $Q$-Criterion: (a) ISO-0.5, (b) LEE-0.5-95, (c) WIN-0.5-95, and (d) ALL-0.5-95, with velocity values colored on the iso-surfaces (the iso-surface corresponds to $Q$=11.8, $Re$=7643, $Ri$=4.00).

## 6. Conclusions

This study investigates the impact of buoyancy flow on the flow field around an isolated cubic building with heated surfaces using LES. The simulation results are validated through wind tunnel tests. Emphasis is placed on low wind speed conditions ranging from 0.5 to 2 m/s. Heating is applied to the windward, leeward, and all walls of the building, resulting in an increase in surface temperature from 20 °C to 95 °C, and the corresponding $Ri$ ranges from 0 to 4.00. The combined numerical simulations and wind tunnel tests uncover general flow structures and the thermal environments.

From the findings, it can be concluded that forced convection plays a dominant role when $Ri$ < 0.33, while buoyancy flow has a minimal effect. The maximum value of $Ri$, which corresponds to the flow characteristics primarily governed by forced convection, is referred to as the critical $Ri$. This critical value is derived from a limited number of cases in this study, and future research could explore a wider array of cases to more accurately identify the critical $Ri$. It is plausible that different regions around the building might exhibit different critical $Ri$ values under varying heating conditions.

Regardless of the specific wall heating cases, the length of the recirculation region exhibits a significant reduction at low Re. When heating all walls, the reduction is most pronounced, with the recirculation zone length shrinking by up to 48.3%, approximately half of the unheated case. Heating the windward wall can increase the length of the roof reattachment zone by at most 8.5%, while heating the leeward wall can lead to a reduction of the reattachment zone length of at most 10.5%. Concerning the vortex core position, as buoyancy effects intensify, there is a general trend of the vortex core positions for both the top and wake vortex moving upward and downstream. Moreover, buoyancy flow can alter the magnitude and direction of the airflow in the far wake. For instance, in the LEE-0.5-95 ($Ri$=4.00) case, within an area approximately 0.25$H_b$ beyond the recirculation zone, the airflow still maintains a noticeable upward trend.

When the leeward wall is heated, most of the heat is carried away vertically by the thermal plume. Upon heating the windward wall, specifically under the $Ri$=4.00 condition, results in the majority of the heat being entrained into the near wake through the recirculation region, causing a rise in temperature on the leeward side. When all the walls are heated at $Ri$=1.34, the buoyancy flow perturbs wake vortex, but the top vortex remains unaffected. At $Ri$ = 2.94 and 4.00, both the wake and top vortex structures become disrupted. Additionally, when the windward wall and all walls are heated, particularly under the $Ri = 4.00$ condition, the thermal plume near the windward wall is strong enough to counteract the downward inertial force, resulting in visible upward motion of the thermal plume. Nevertheless, near the heated wall surface, temperature drops swiftly, reaching about 30% of the heating temperature at a distance of $0.05H_b$ from the wall.

At low $Re$ values, in the cases of no heating and when heating the windward wall, the position of $k_{max}$ is centrally located on roof. However, with heating of the leeward wall, $k_{max}$ appears at the trailing edge of the roof. When all walls are heated, $k_{max}$ is located near the ground on the leeward side, a result of the combined effects of buoyancy fluctuation and shear force. In most of the investigated cases, the location of the maximum absolute value of $<U'W'>/U_{ref}^2$ coincides with the location of $k_{max}$. Furthermore, under low $Re$ and high surface temperatures, the thickness of the free shear layer widens due to the diffusion effect. When considering the same incoming wind speed and heating temperature, the widening degree of the free shear layer follows the order: heating all walls, heating the windward wall, and heating the leeward wall.

In conclusion, this study of the influence of thermal effects on flow field characteristics offers potential integration with research on architectural planning and pollutant dispersion. For instance, while a high $k$ value promotes air mixing and heat transfer, it also simultaneously results in an elevated rate of pollutant diffusion within buildings. An optimally designed flow field can better facilitate fresh air intake and pollutant expulsion, thus improving urban air quality and reducing pollutant buildup. This research sheds light on the dynamics of airflow and temperature distribution in outdoor environments. It underscores the importance of considering the interplay between incoming wind speed and buoyancy-induced flow. Emphasis is especially placed on the building's corners and wall areas, as these regions could play a central role in the accumulation and distribution of thermal airflows at pedestrian level. While this study presents an idealized context, real-world scenarios are likely to be influenced by neighboring buildings. This aspect warrants further exploration in future research.

## Declaration of Competing Interest

The authors declare that they have no known competing financial interests or personal relationships that could have appeared to influence the work reported in this paper.

## Acknowledgments

This research was supported by the National Natural Science Foundation of China under the project reference No. 52078353, the Research Scheme of Research Grants Council of Hong Kong SAR, China (Project No. T22-504/21R), International Cooperation project of Science and Technology Commission of Shanghai Municipality (No. 22200711400), and the Fundamental Research Funds for the Central Universities.

## Appendix A. Comparative analysis of normalized time-averaged $U$x along various measurement lines.

Under different cases, the normalize time-averaged $U_x$ along the measurement line $x/H_b$=-0.25

on the windward side shows minimal variation, as depicted in Fig. A. However, at $Re$=7643, significant disparities in the time-averaged velocity are observed on the roof ($x/H_b$=0.5) and the leeward side of the building ($x/H_b$=1.5 and 3.5) under different heating conditions. Specifically, on the measurement line at $x/H_b$=3.5 (beyond the recirculation region), the buoyancy effect becomes more prominent as the $Ri$ increases, resulting in higher time-averaged velocities below the building height. In other words, the action of buoyancy flow can influence the magnitude and direction of airflow in the far wake region. This phenomenon becomes less apparent as the $Re$ increases. When $Re$=50955, the time-averaged velocities under different cases on the four measurement lines tend to converge and closely align with the non-heated condition. At this point, forced convection dominates, and the influence of buoyancy flow is minimal.

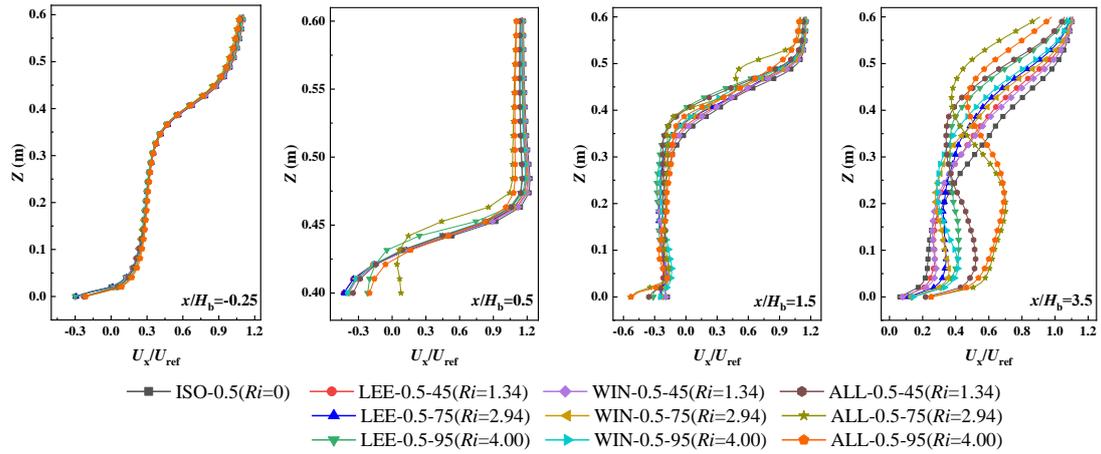

$Re$=7643, $U_{ref}$=0.5m/s

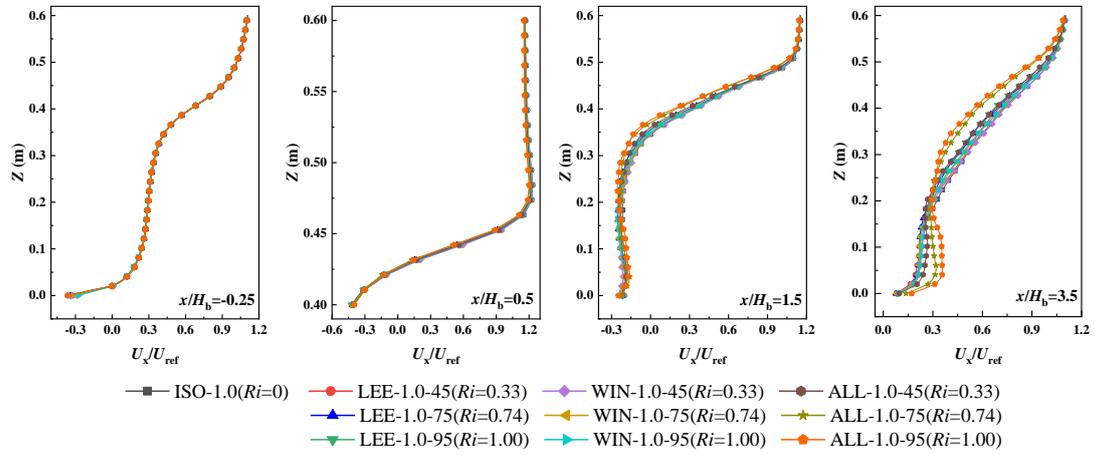

$Re$=25478, $U_{ref}$=1.0m/s

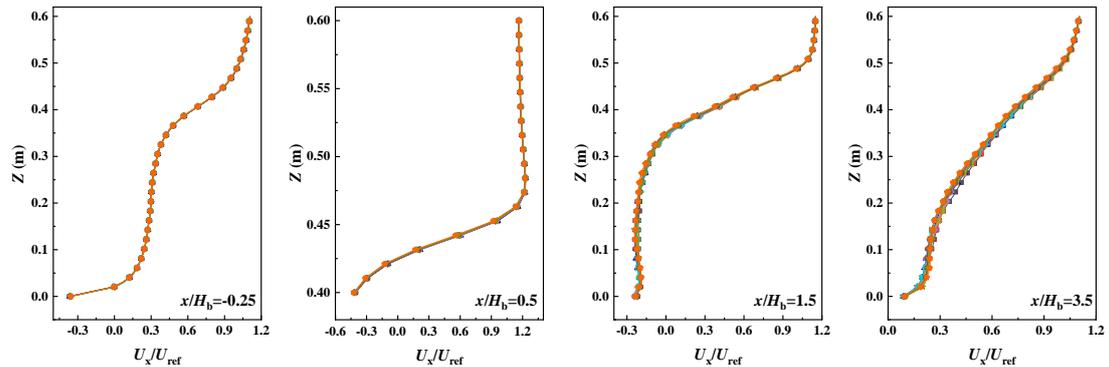

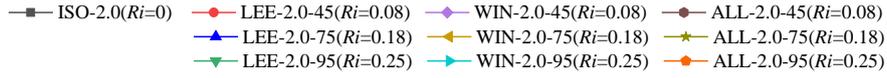

$Re$=50955, $U_{ref}$=2.0m/s

**Fig. A.** The normalized time-averaged $U_x$ at $x/H_b$=-0.25, 0.5, 1.5 and 3.5 for different reference wind speeds.

## Appendix B. Time-averaged velocity and temperature fields when heating the leeward wall ($Re$=50955).

When $Re$=50955 ($Ri$<0.33), the phenomenon of thermal airflow transferring to the roof and leeward side nearly disappears. In this situation where forced convection dominates, the buoyancy effect is not very pronounced.

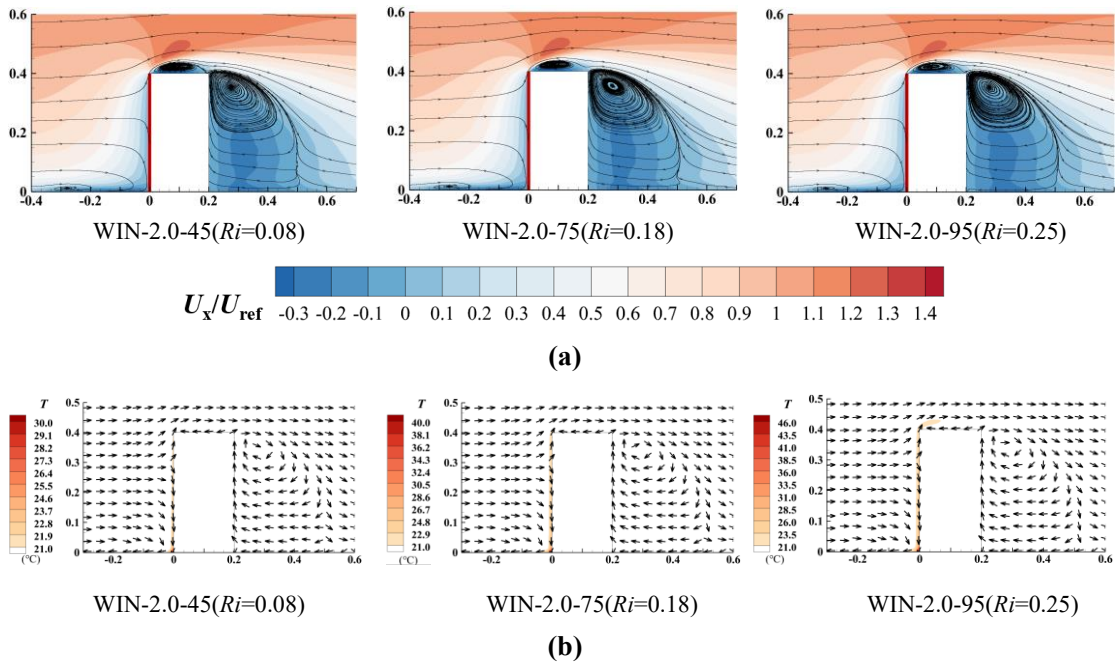

**Fig. B.** (a)Time-averaged airflow streamlines colored by the normalized time-averaged $U_x$ and (b) time-averaged velocity vectors and temperature over the vertical central plane at $y/H_b$=0.5 (the heated wall is represented by a red line; the dimensions are in [m]).